\documentclass[a4paper,11pt]{article}
\pdfoutput=1 

\usepackage{jheppub} 
\usepackage[bottom]{footmisc}
\usepackage{amssymb}
\usepackage{amsmath}
\usepackage{amsthm}
\usepackage[usenames,dvipsnames]{xcolor}
\usepackage{epsfig}
\usepackage{dcolumn}
\usepackage{tikz}
\usetikzlibrary{shapes.geometric, arrows,positioning}
\usetikzlibrary{calc,decorations.pathreplacing}
\usepackage{upgreek}
\usepackage{setspace}
\usepackage{array,multirow,bigdelim,arydshln}
\usepackage{appendix}
\usepackage{xparse}
\usepackage{stmaryrd}
\usepackage[T1]{fontenc} 
\usepackage{mathtools}
\usepackage{physics} 
\usepackage{adjustbox}
\usepackage{multirow}
\usepackage{graphicx} 
\usepackage{subcaption}
\usepackage{float} 
\graphicspath{{./images/}}
\usepackage{comment}
\usepackage[nottoc]{tocbibind}
\usepackage{hyperref}
\usepackage[utf8]{inputenc}
\usepackage{CJK}
\hypersetup{
	colorlinks,
	urlcolor=Maroon,
	linkcolor=Maroon,
	citecolor=Maroon
	}

\NewDocumentCommand{\binomial}{omm}
 {%
  \genfrac(){0pt}{}{#2}{#3}%
  \IfValueT{#1}{_{\!#1}}%
 }
\NewDocumentCommand{\eulerian}{omm}
 {%
  \genfrac<>{0pt}{}{#2}{#3}%
  \IfValueT{#1}{_{\!#1}}%
 }
\newcommand{\ab}[2]{\langle #1\,#2\rangle}
\newcommand{\sqb}[2]{[#1\,#2]}

\usepackage{latexsym}

\theoremstyle{plain}

\theoremstyle{definition}

\newcommand{\ve}{\varepsilon}
\newcommand{\e}{\epsilon}

\def\bea#1\eea{\begin{eqnarray}#1\end{eqnarray}}
\def\be#1\ee{\begin{equation}#1\end{equation}}
\def\ee{\end{equation}}
\def\ba#1\ea{\begin{align}#1\end{align}}

\usepackage{amsmath}
\usepackage{multicol}
\usepackage{bbm}
\usepackage{enumerate}

\usepackage{amsthm}
\usepackage{mathrsfs}
\usepackage{upgreek}
\usepackage{amssymb}
\usepackage{bm}
\usepackage{setspace}
\usepackage{array,multirow,arydshln}
\usepackage{bigdelim}
\usepackage{scalerel}
\usepackage{diagbox}

\usepackage{tabularx}

\usepackage{tikz}

\usetikzlibrary{shapes.geometric,arrows,arrows.meta,decorations.pathmorphing,decorations.markings,patterns}

\def\<{\langle}
\def\>{\rangle}

\def\e{\epsilon}

\def\r{\ref}

\usepackage[percent]{overpic}
\usepackage{multirow} 
\usepackage{slashed}

\def\Sc{{\cal S}}
\def\lf{\left}
\def\ri{\right}

\title{Subleading Collinear Limits of Yang--Mills Amplitudes from Recursions}

\author[]{Jin Dong}
\author[]{and Stephan Stieberger}

\affiliation[]{Max-Planck-Institut f\"ur Physik, Werner-Heisenberg-Institut, Boltzmannstr. 8,\\[2mm] 85748 Garching bei M\"unchen, Germany}

\emailAdd{jindong@mpp.mpg.de}
\emailAdd{stephan.stieberger@mpp.mpg.de}

\abstract{We study the subleading adjacent-collinear limit of tree-level color-ordered Yang--Mills amplitudes. In general dimensions, a Lorentz null rotation transports the polarizations of collinear legs along the collinear path while preserving transversality and gauge equivalence.  The finite coefficient then separates into a pole-free hard contribution, computed by a channel-deleted Berends--Giele recursion, and an explicit contribution from the adjacent factorization channel.  In four dimensions, we construct the same coefficient directly from a collinear-aware Britto-Cachazo-Feng-Witten recursion whose terminal data are closed MHV and $\overline{\rm MHV}$ formulas.  Both constructions apply at arbitrary momentum fraction and extend to collinear gluons with distinct polarizations; at the symmetric split, the auxiliary-vector dependence cancels for equal polarizations and for the polarization-symmetrized mixed limit.  Open--closed disk relations then give a direct construction of one-graviton Einstein--Yang--Mills amplitudes and their higher-derivative string corrections from the recursively generated coefficients.  A self-contained \textsc{Mathematica} implementation accompanies the paper.}

\preprint{MPP-2026-145}
\begin{document}

\begin{CJK*}{UTF8}{}
\CJKfamily{gbsn}
\maketitle
\end{CJK*}
\addtocontents{toc}{\protect\setcounter{tocdepth}{2}}

		\tikzset{
		particles/.style={dashed, postaction={decorate},
			decoration={markings,mark=at position .5 with {\arrow[scale=1.5]{>}}
		}}
	}
	\tikzset{
		particle/.style={draw=black, postaction={decorate},
			decoration={markings,mark=at position .5 with {\arrow[scale=1.1]{>}}
		}}
	}
	\def  \layersep {.6cm}

\section{Introduction}
\label{sec:introduction}
Factorization in infrared limits is one of the fundamental analytic properties of gauge-theory amplitudes. When two massless external momenta become parallel, an intermediate state goes on shell and the amplitude develops a collinear singularity. For a color-ordered Yang--Mills tree amplitude, this singularity occurs only when the two legs are adjacent. Its leading behavior takes the universal form~\cite{Mangano:1990by,Dixon:1996wi,Elvang:2013cua, Badger:2023eqz}
\begin{equation}
 A_n^{\rm tree}(\ldots,a^{h_a},b^{h_b},\ldots)
 \xrightarrow{\,a\parallel b\,}
 \sum_{h_P}
 {\rm Split}_{-h_P}^{\rm tree}(a^{h_a},b^{h_b};x)\,
 A_{n-1}^{\rm tree}(\ldots,P^{h_P},\ldots),
\label{eq:intro-leading-factorization}
\end{equation}
where $p_a\to xP$, $p_b\to(1-x)P$, and the splitting amplitude depends only on the two collinear states and their momentum fraction. All dependence on the remaining hard particles resides in the lower-point amplitude. This separation of scales underlies the use of collinear limits as rigorous checks on amplitude constructions and supplies the local counterterms required in perturbative QCD calculations~\cite{Catani:1996jh,Catani:1996vz}. The singular factorization also persists beyond tree level, with universal loop splitting amplitudes multiplying lower-loop hard amplitudes~\cite{Bern:1994zx,Kosower:1999rx,Kosower:1999xi}.

The universality of Eq.~\eqref{eq:intro-leading-factorization} concerns only the singular term. Considerably less is known about the next term in the expansion. At tree level this contribution is finite and is not determined by a splitting function multiplying a lower-point amplitude: it receives one contribution from expanding the singular factorization channel and another from diagrams that are regular in that channel. Moreover, its explicit form depends on the path used to approach collinear kinematics, including the continuation of the external polarizations. Thus a definition of the subleading coefficient requires more information than the limiting momenta $p_a=xP$ and $p_b=(1-x)P$ alone.

Several results nevertheless indicate that this finite sector has a nontrivial structure. One of the authors and Taylor showed that weighted combinations of subleading collinear Yang--Mills amplitudes are related to Einstein--Yang--Mills amplitudes in which the two collinear gluons are replaced by one graviton~\cite{Stieberger:2014hba,Stieberger:2015kia_PLB}. A complementary analysis based on the Cachazo--He--Yuan formulae~\cite{Cachazo:2013gna, Cachazo:2013hca,Cachazo:2013iea} expressed the subleading limit as a lower-point integrand convolved with a universal collinear kernel and extended the construction to gravitons and scalar effective field theories~\cite{Nandan:2016CHY}. These representations expose important universal ingredients, but they do not by themselves provide a direct amplitude-level recursion for each individual finite coefficient. In particular, a systematic construction valid at arbitrary momentum fraction, in general dimensions, and for both equal and distinct polarizations has remained unavailable.

This paper develops direct recursive constructions for that finite term.  We consider adjacent legs $n-1$ and $n$, introduce a parameter $\epsilon$ that measures their departure from the collinear surface, and denote the complete $\epsilon^0$ coefficient the \emph{strict subleading collinear limit}.  Once the momentum path and the transport of external polarizations have been specified, this coefficient is unambiguous. Our aim is to compute it without first constructing a full amplitude with $\epsilon$-dependent kinematics and then performing a Laurent expansion.

In general dimensions, we transport the collinear momentum and its polarization by the same Lorentz null rotation. The resulting polarizations remain null and transverse. This gives a covariant collinear path for arbitrary momentum fraction $x$.  The strict coefficient naturally separates into two pieces: a hard term obtained by deleting the adjacent two-particle current from Berends--Giele (BG) recursion~\cite{Berends:1987me}, and a finite term obtained by expanding the deleted factorization block.  The latter is essential at generic $x$, whereas for equal polarizations it
vanishes at the symmetric split $x=1/2$.  

Four dimensions admit a complementary on-shell construction.  Using spinor-helicity variables and Britto-Cachazo-Feng-Witten (BCFW) recursion \cite{Britto:2004ap,Britto:2005fq}, we classify recursive terms according to whether the two collinear legs occur in different subamplitudes or in the same one.  The first class is evaluated directly on the collinear surface; the second inherits the collinear operation recursively.  Closed MHV and $\overline{\rm MHV}$ formulas terminate the recursion.  This algorithm acts on the recursive representation itself and never expands the complete amplitude in $\epsilon$.

We also treat collinear gluons with distinct polarizations.  In four dimensions these are the ordered mixed-helicity configurations $(-,+)$ and $(+,-)$.  Each ordered coefficient can depend on the auxiliary reference at $x=1/2$, but the dependence cancels in their polarization-symmetrized sum. The covariant construction makes the same cancellation manifest as the antisymmetry of the finite factorization contribution.

The recursive results have a second application through mixed open--closed disk amplitudes.  Monodromy relations replace a closed-string insertion by a weighted sum over two additional open-string insertions \cite{Stieberger:0907,Stieberger:2015kia}.  In the field-theory collinear limit these become two gluons representing a graviton \cite{Stieberger:2014hba,Stieberger:2015kia_PLB,Stieberger:2016lag}.  A companion paper \cite{DongStieberger:toappear} uses this relation to organize strict Yang--Mills coefficients in terms of one-graviton Einstein--Yang--Mills amplitudes, higher-derivative corrections, and homogeneous BCJ-like relations.  For the present purpose, once the Yang--Mills coefficients have been generated recursively, the disk relation and its kernels directly construct the EYM amplitudes and their string corrections.

The ancillary \textsc{Mathematica} package implements both recursions, their equal- and mixed-polarization variants, and the reduction of covariant expressions to four-dimensional helicity components.  The implementation is analytic; independent numerical tree-amplitude tools are used only for checks.

The paper is organized as follows. Section~\ref{sec:general-dimensional-collinear} develops the covariant collinear path, derives the general-dimensional result, and introduces the modified Berends--Giele recursion for its hard contribution. Section~\ref{sec:four-dimensional-collinear} gives the four-dimensional BCFW construction and explicit examples. Section~\ref{sec:distinct-polarizations} extends both formulations to distinct polarizations. Section~\ref{sec:eym-from-collinear} describes the construction of EYM amplitudes and string corrections from collinear Yang--Mills data, and Section~\ref{sec:mathematica-package} documents the accompanying implementation. Appendix~\ref{app:four-dimensional-reduction} shows explicitly how the covariant formulas reduce to the familiar four-dimensional helicity splitting amplitudes, while Appendix~\ref{app:modified-bg} proves the consistency of the modified Berends--Giele recursion.

\paragraph{Note added.}
The Codex AI agent, running the GPT-5.6 Sol model, was used at several stages in the preparation of this work.  In particular, the accompanying Mathematica package was developed with Codex from source code written by the authors and from publicly available packages.  The BCFW-recursion module draws in part on the computational tools of Ref.~\cite{Bourjaily:2023uln}. All analytical results and computer-generated outputs were checked by the authors.

\section{Subleading collinear limits in general dimensions from modified BG recursion}
\label{sec:general-dimensional-collinear}

We approach the adjacent collinear surface with the standard on-shell parametrization~\cite{Catani:1996vz}
\begin{equation} \label{eq: collinear momenta}
\begin{aligned}
p_{n-1}^\mu &= x\, P^\mu
- \epsilon\, k_\perp^\mu
- \frac{\epsilon^2\, k_\perp^2}{2 x\, P\!\cdot K}\, K^\mu , \\[6pt]
p_n^\mu &= (1-x)\, P^\mu
+ \epsilon\, k_\perp^\mu
- \frac{\epsilon^2\, k_\perp^2}{2(1-x)\, P\!\cdot K}\, K^\mu .
\end{aligned}
\end{equation}
Here the ``parent'' momentum $P$, whose momentum is distributed between the
two collinear legs according to the momentum fraction $x$ with $0\leq x \leq1$, and the reference
momentum $K$ are both null,
\begin{equation}
P^2=K^2=0,
\qquad
P\cdot K\neq 0,
\end{equation}
while $k_\perp$ is transverse to both null directions,
\begin{equation}
k_\perp \!\cdot P = k_\perp \!\cdot K = 0 .
\end{equation}
Both ``daughter'' momenta remain exactly on shell, while their pair invariant is
\begin{equation} \label{eq: divergent pole}
p_{n-1}^2 = p_n^2 = 0,
\qquad
s_{n-1,n} \equiv (p_{n-1}+p_n)^2
= -\frac{\epsilon^2\, k_\perp^2}{x(1-x)} .
\end{equation}

We next construct the daughter polarizations.  Let $\varepsilon_P^\mu$ denote a complex null polarization associated with the parent momentum $P^\mu$, satisfying
\begin{equation}
 \varepsilon_P \cdot P=0,
 \qquad
 \varepsilon_P^2=0.
 \label{eq:null-parent-polarization}
\end{equation}
For simplicity, we take the two collinear legs to share the same parent
polarization. The case of distinct polarizations is discussed in
Section~\ref{sec:distinct-polarizations}. To continue the polarization away from the strict collinear configuration,
it is not sufficient simply to use $\varepsilon_{n-1}=\varepsilon_n=\varepsilon_P$. For a generic state, one would then have
$p_{n-1}\cdot\varepsilon_P,p_n\cdot\varepsilon_P=O(\epsilon)$, so the
polarizations would no longer remain transverse to the deformed momenta.
One could attempt to restore transversality by adding a correction
proportional to the reference momentum $K$. However, imposing
$p_i\cdot\varepsilon_i=0$ in this way does not in general preserve the null
condition, and one may generically obtain $\varepsilon_i^2\neq0$. A more
systematic prescription is instead to transform each momentum and its
polarization by the same Lorentz transformation, thereby preserving all
their inner products automatically.

For any vector \(a^\mu\) transverse to both null directions,
\begin{equation}
    a\cdot P=a\cdot K=0,
\end{equation}
define the Lorentz-algebra generator
\begin{equation}
    \Omega(a)^\mu{}_\nu
    :=\frac{a^\mu K_\nu-K^\mu a_\nu}{P\cdot K}.
\end{equation}
Since \(K^2=a\cdot K=0\), it is nilpotent of degree three,
\begin{equation}
    \Omega(a)^3=0.
\end{equation}
The associated finite null rotation is
\begin{equation}
\begin{aligned}
    {\cal R}(a)^\mu{}_\nu
    &:=\big(e^{\Omega(a)}\big)^\mu{}_\nu ,
    \\
    \big({\cal R}(a)v\big)^\mu
    &:={\cal R}(a)^\mu{}_\nu v^\nu
    =v^\mu
    +\frac{K\cdot v}{P\cdot K}a^\mu
    -\frac{a\cdot v}{P\cdot K}K^\mu
    -\frac{a^2(K\cdot v)}{2(P\cdot K)^2}K^\mu .
\end{aligned}
\label{eq:null-rotation}
\end{equation}
The name ``null rotation'' refers to the fact that this Lorentz transformation fixes the null vector $K$.  In particular,
\begin{equation}
 {\cal R}(a)K=K,
 \qquad
 {\cal R}(a)P=P+a-\frac{a^2}{2P\cdot K}K,
 \qquad
 {\cal R}(a)u\cdot{\cal R}(a)v=u\cdot v.
 \label{eq:null-rotation-properties}
\end{equation}
Consequently, the momenta in Eq.~\eqref{eq: collinear momenta} and their polarizations can be written as
\begin{equation}
\begin{aligned}
 p_{n-1}&=x\,{\cal R}\!\left(-\frac{\epsilon}{x}k_\perp\right)P,
 &\qquad
 \varepsilon_{n-1}
 &={\cal R}\!\left(-\frac{\epsilon}{x}k_\perp\right)\varepsilon_P,
 \\
 p_n&=(1-x)\,{\cal R}\!\left(\frac{\epsilon}{1-x}k_\perp\right)P,
 &
 \varepsilon_n
 &={\cal R}\!\left(\frac{\epsilon}{1-x}k_\perp\right)\varepsilon_P.
\end{aligned}
\label{eq:pol_projection}
\end{equation}
Explicitly,
\begin{align} \label{eq: collinear pol}
\varepsilon_{n-1}^\mu
&=\varepsilon_P^\mu
-\epsilon\frac{\varepsilon_P\cdot K}{xP\cdot K}k_\perp^\mu
+\epsilon\frac{\varepsilon_P\cdot k_\perp}{xP\cdot K}K^\mu
-\epsilon^2\frac{k_\perp^2(\varepsilon_P\cdot K)}
 {2x^2(P\cdot K)^2}K^\mu,
\\[6pt]
\varepsilon_{n}^\mu
&=\varepsilon_P^\mu
+\epsilon\frac{\varepsilon_P\cdot K}{(1-x)P\cdot K}k_\perp^\mu
-\epsilon\frac{\varepsilon_P\cdot k_\perp}{(1-x)P\cdot K}K^\mu
-\epsilon^2\frac{k_\perp^2(\varepsilon_P\cdot K)}
 {2(1-x)^2(P\cdot K)^2}K^\mu.
\end{align}

This construction preserves transversality and the null norm exactly,
\begin{equation}
 p_i\cdot\varepsilon_i=0,
 \qquad \varepsilon_i^2=0 \,,
\end{equation}
which follows immediately from the last identity in Eq.~\eqref{eq:null-rotation-properties}.

The null rotation is a Lorentz transformation, not a gauge transformation. Its relevance to gauge invariance is that it is well defined on the parent gauge orbit.  If $p_i=z_i{\cal R}(a_i)P$, with $z_i=x$ or $1-x$, then
\begin{equation}
 {\cal R}(a_i)(\varepsilon_P+\alpha P)
 ={\cal R}(a_i)\varepsilon_P+\alpha{\cal R}(a_i)P
 =\varepsilon_i+\frac{\alpha}{z_i}p_i.
 \label{eq:null-rotation-gauge-orbit}
\end{equation}
Thus two gauge-equivalent representatives of the parent polarization are transported to gauge-equivalent representatives of each daughter polarization.  The construction therefore defines a map between the physical polarization spaces along the entire path.  The same prescription therefore applies when the two collinear legs are transported from two independent parent polarizations, as discussed in Section~\ref{sec:distinct-polarizations}. In four dimensions, changing the reference spinor of $\varepsilon_P$ shifts it by a multiple of $P$; Eq.~\eqref{eq:null-rotation-gauge-orbit} shows directly that this becomes a gauge shift of the corresponding daughter polarization.  Reference-spinor independence is therefore built into the continuation.

Throughout the paper we expand an adjacent color-ordered amplitude as
\begin{equation}\label{eq:coll-notation}
A_n(\epsilon)=\sum_{r=-1}^{\infty}A_n\Big|^{(r)}_{\rm coll.},
\end{equation}
so $A_n\big|^{(r)}_{\rm coll.}$ denotes the complete term at order $\epsilon^r$. For a specified ordering we use the equivalent shorthand for the subleading finite order
\begin{equation}\label{eq:sub-notation}
A_n(1,\rho,n-1,n)_{\rm sub}:
=A_n(1,\rho,n-1,n)\Big|^{(0)}_{\rm coll.}.
\end{equation}
In what follows, we focus primarily on the canonical ordering; other orderings are obtained by relabeling.

\subsection{Leading collinear limit}
Consider the color-ordered amplitude $A(1,2,\ldots,n)\equiv A_n$.  The only propagator singular on the collinear limit is $1/s_{n-1,n}=O(\epsilon^{-2})$, so the singular contribution is entirely carried by the adjacent factorization channel,
\begin{equation} \label{eq: leading collinear}
    \frac{1}{s_{n-1,n}}\mathrm{Res}_{s_{n-1,n}=0} A_n= \frac{1}{s_{n-1,n}} \sum_{\rm states} A(1,2,\ldots,n-2,I) A({-}I,n-1,n) \,
\end{equation}
where the internal-state sum can be represented as
\begin{equation}
\sum_{\rm states} \varepsilon_I^\mu\ \varepsilon_{-I}^\nu
= \eta^{\mu \nu}
- \frac{p_I^\mu q^\nu + p_I^\nu q^\mu}{p_I \cdot q}\,.
\end{equation}
The internal momentum is $I=p_{n-1}+p_n$.  On the factorization surface $I^2=0$, both subamplitudes obey the Ward identity with respect to $I$; hence the $q$-dependent part of the polarization sum drops out.  In the strict collinear limit, $I$ is identified with the null parent momentum $P$.

It is convenient to strip the internal polarization from the three-point factor and write
\begin{equation}
    A(-I,n-1,n)=\varepsilon_{-I,\mu}\,V^\mu(-I,n-1,n),
\end{equation}
with
\begin{equation}
\begin{aligned}
V^\mu(-I,n-1,n)={}&-2\varepsilon_{n-1}^\mu\,
p_{n-1}\!\cdot\!\varepsilon_n
+2\varepsilon_n^\mu\,p_n\!\cdot\!\varepsilon_{n-1}
\\
&+(\varepsilon_{n-1}\!\cdot\!\varepsilon_n)
(p_{n-1}-p_n)^\mu\,.
\end{aligned}
\label{eq:oriented-three-current}
\end{equation}
Here all momenta are outgoing.  This convention is consistent with the color-ordered Berends--Giele vertex and with the Parke--Taylor normalization used below. We expand the 3-point amplitude in small $\epsilon$ along the null-rotation path.  
\begin{equation} \label{eq: 3-pt collinear}
\begin{aligned}
    A({-}I,n-1,n)\Big|_{\rm coll.}
    &= \e \,\frac{2\ve_{-I}\cdot \ve_P\,
    \ve_P \cdot k_\perp}{x(1-x)}
    \\
    &\quad+\e^2\,
    \frac{(2x-1)(\ve_P\cdot K)
    (\ve_{-I}\cdot\ve_P)k_\perp^2}
    {x^2(1-x)^2P\cdot K}
    +O(\epsilon^3).
\end{aligned}
\end{equation}
Combining Eqs.~\eqref{eq: divergent pole} and \eqref{eq: 3-pt collinear} with Eq.~\eqref{eq: leading collinear} gives
\begin{equation}
   A_n \Big|^{(-1)}_{\rm coll.}= -\frac{2\ve_P \cdot k_\perp } {\epsilon\, k_\perp^2}  \ \varepsilon_P^\mu A_\mu(1,2,\ldots,n-2,I)\,,
\end{equation}
where
\begin{equation}
    A(1,2,\ldots,n-2,I)=\varepsilon_I^\mu A_\mu(1,2,\ldots,n-2,I)\,.
\end{equation}
At this stage the lower-point factor still carries the slightly off-shell leg $I$.  Its difference from the null vector $P$ starts only at $O(\epsilon^2)$:
\begin{equation}
    p_I^\mu=P^\mu+\delta p^\mu,\qquad \delta p^\mu=\mathcal{O}(\epsilon^2)\,.
\end{equation}
Accordingly,
\begin{equation}
    A_\mu(1,2,\ldots,n-2,I)=A_\mu(1,2,\ldots,n-2,P)+\mathcal{O}(\epsilon^2)\,,
\end{equation}
so after multiplication by the three-point factor and the pole $1/s_{n-1,n}$ the induced correction is beyond subleading order:
\[
\frac{1}{s_{n-1,n}}\times \mathcal{O}(\epsilon^2)\times \mathcal{O}(\epsilon)
\sim \mathcal{O}(\epsilon)\,.
\]
Thus $I$ may be replaced by $P$ through subleading order, and
\begin{equation}
   A_n \Big|^{(-1)}_{\rm coll.}= -\frac{2\ve_P \cdot k_\perp } {\epsilon\, k_\perp^2}   \  A(1,2,\ldots,n-2,P)\, .
\end{equation}
Its reduction to the familiar four-dimensional same-helicity splitting amplitudes is displayed explicitly in Appendix~\ref{app:four-dimensional-reduction}.
\subsection{Subleading collinear limit}
The subleading collinear term is the coefficient at $O(\epsilon^0)$.  It receives contributions both from the expansion of the adjacent pole and from terms regular in that channel.  We therefore write
\begin{equation} \label{eq: decomposition}
    A_n= \left(A_n-\frac{1}{s_{n-1,n}}\mathrm{Res}_{s_{n-1,n}=0} A_n\right)+ \frac{1}{s_{n-1,n}}\mathrm{Res}_{s_{n-1,n}=0} A_n \,.
\end{equation}
For the second term, the $\epsilon^2$ coefficient in Eq.~\eqref{eq: 3-pt collinear} gives
\begin{equation}
\left(\frac{1}{s_{n-1,n}}
\mathrm{Res}_{s_{n-1,n}=0} A_n \right)
\Big|^{(0)}_{\rm coll.}
=\frac{(1-2x)\varepsilon_P\cdot K}
{x(1-x)P\cdot K}A(1,2,\ldots,n-2,P).
\label{eq:equal-pole-finite}
\end{equation}
The first term in Eq.~\eqref{eq: decomposition} is regular in $s_{n-1,n}$, so its $O(\epsilon^0)$ coefficient is obtained by the strict replacement $p_{n-1} \to xP,\ p_n \to (1-x)P$ and $\varepsilon_{n-1},\varepsilon_n \to \varepsilon_P$, with no further expansion.  The result is

\begin{align}  \label{eq: subleading}
    A_n\Big|^{(0)}_{\rm coll.}=& \left(A_n-\frac{1}{s_{n-1,n}}\mathrm{Res}_{s_{n-1,n}=0} A_n\right)\Bigg|_{\substack{\hspace{-0.3cm}  p_{n-1} \to xP \\  p_n \to (1-x)P  \\
    \varepsilon_{n-1},\varepsilon_n \to \varepsilon_P}  
    }
    +\frac{(1-2x)\varepsilon_P\cdot K}
    {x(1-x)P\cdot K}A(1,2,\ldots,n-2,P).
\end{align}
Remarkably, while the lower-point amplitude in the second term can itself be computed using Berends--Giele recursion, the hard contribution in Eq.~\eqref{eq: subleading} can be evaluated directly using the modified BG recursion introduced in Section~\ref{subsec:modified-bg}, without first constructing the residue and subtracting it from the full amplitude. The subtle point in this construction is that a numerator factor can cancel the deleted two-particle propagator and apparently leave a finite term. We explain below how the modified current is defined and prove that no strict hard contribution is lost in Appendix~\ref{app:modified-bg}.

We emphasize that the hard term contains neither $K^\mu$ nor
$k_\perp^\mu$. Moreover, the explicit pole correction is proportional to
$1-2x$ and therefore \textbf{vanishes} at $x=1/2$. Consequently, the
\textbf{independence} of the symmetric strict coefficient from both auxiliary vectors
is manifest.

Before concluding this subsection, let us discuss the singularity structure. The singularities of color-ordered amplitudes are conveniently described by the planar variables~\cite{Arkani-Hamed:2017mur}
\begin{equation}
    X_{i,j}:=(p_i+p_{i+1}+\ldots+p_{j-1})^2\,,
\end{equation}
with $X_{i,j}=X_{j,i}$ and $X_{i,i}=X_{i,i+1}=0$.  For $i<j$, each nontrivial $X_{i,j}$ labels a planar channel of $A(1,2,\ldots,n)$. There are $n(n-3)/2$ such variables, which form a natural basis for the space of Lorentz products of external momenta.  Under the strict replacement $p_{n-1}\to xP$ and $p_n\to(1-x)P$, they obey
\begin{equation} \label{eq: old pole}
    X_{i,j} \to X_{i,j}, \quad j\neq n \quad {\rm and}\quad X_{1,n-1}\to 0\,,
\end{equation}
\begin{equation} \label{eq: new pole}
    X_{i,n} \to x X_{1,i}+(1-x)X_{i,n-1}\,.
\end{equation}
Thus, besides ordinary $(n{-}1)$-point channels, the hard term contains poles on the linear combinations in Eq.~\eqref{eq: new pole}.  The endpoint cases $i=2$ and $i=n-2$ specialize to
\begin{equation} \label{eq: new special pole}
X_{2,n} \to (1-x)X_{2,n-1},
\qquad
X_{n-2,n} \to x X_{1,n-2}.
\end{equation}
Double poles in $X_{2,n-1}$ and $X_{1,n-2}$ may therefore arise from the coincidence of an ordinary lower-point channel with the corresponding specialization of Eq.~\eqref{eq: new pole}.

\subsection{Modified Berends--Giele recursion for the hard contribution}
\label{subsec:modified-bg}

The hard term in Eq.~\eqref{eq: subleading} can be evaluated by a simple modification of the color-ordered Berends--Giele recursion~\cite{Berends:1987me}. For a contiguous ordered word $\alpha$, let $p_\alpha:=\sum_{a\in\alpha}p_a$ and $s_\alpha:=p_\alpha^2$. In the conventions of this paper, the ordered cubic and quartic contractions are
\begin{equation}
\begin{aligned}
 V_3^\mu(p,q;J_1,J_2)={}&
 (J_1\!\cdot J_2)(p-q)^\mu
 +\big[J_1\!\cdot(p+2q)\big]J_2^\mu
 -\big[J_2\!\cdot(2p+q)\big]J_1^\mu,
 \\
 V_4^\mu(J_1,J_2,J_3)={}&
 2(J_1\!\cdot J_3)J_2^\mu
 -(J_2\!\cdot J_3)J_1^\mu
 -(J_1\!\cdot J_2)J_3^\mu.
\end{aligned}
\label{eq:bg-vertices}
\end{equation}
Starting from $J^\mu(i)=\varepsilon_i^\mu$, the ordinary current obeys
\begin{equation}
\begin{aligned}
 J^\mu(\alpha)=\frac{1}{s_\alpha}\Bigg[&
 \sum_{\alpha=\beta\gamma}
 V_3^\mu\big(p_\beta,p_\gamma;J(\beta),J(\gamma)\big)+\sum_{\alpha=\beta\gamma\delta}
 V_4^\mu\big(J(\beta),J(\gamma),J(\delta)\big)
 \Bigg],
\end{aligned}
\label{eq:bg-current}
\end{equation}
where the sums run over nonempty ordered deconcatenations. Writing $\mathcal N^\mu(\alpha):=s_\alpha J^\mu(\alpha)$ for the amputated current, the usual color-ordered amplitude is
\begin{equation}
 A_n(1,2,\ldots,n)
 =\varepsilon_{1,\mu}\,\mathcal N^\mu(2,3,\ldots,n).
\label{eq:bg-amplitude}
\end{equation}

The modified current satisfies the same recursion \eqref{eq:bg-current}, with $J$ replaced everywhere by $J_{\rm mod}$, but obeys the boundary conditions
\begin{equation}
 J_{\rm mod}^\mu(i):=\varepsilon_i^\mu,
 \qquad
 J_{\rm mod}^\mu(n-1,n):=0.
\label{eq:bg-modified-boundary}
\end{equation}
Thus every graph containing the special two-particle subcurrent is removed at its source. If $\mathcal N_{\rm mod}^\mu(\alpha):=s_\alpha J_{\rm mod}^\mu(\alpha)$, the associated amputated amplitude is
\begin{equation}
 A_n^{\rm mod}(1,2,\ldots,n)
 :=\varepsilon_{1,\mu}\,
 \mathcal N_{\rm mod}^\mu(2,3,\ldots,n).
\label{eq:bg-modified-amplitude}
\end{equation}
Its strict value gives the hard contribution,
\begin{equation}
 \mathcal H_n(\varepsilon_P,\varepsilon'_P)
 =A_n^{\rm mod}\bigg|_{\substack{
 p_{n-1}=xP,\ p_n=(1-x)P\\
 \varepsilon_{n-1}=\varepsilon_P,\ 
 \varepsilon_n=\varepsilon'_P}},
\label{eq:bg-hard}
\end{equation}
with $\varepsilon'_P=\varepsilon_P$ in the equal-polarization case relevant here. The same definition will be used for distinct polarizations in Section~\ref{sec:distinct-polarizations}.

There is an important subtlety. Although the pair current carries the propagator $1/s_{n-1,n}$, a factor of $s_{n-1,n}$ can arise from its contraction with the remainder of a BG graph and cancel that propagator. It is therefore not legitimate to identify channel deletion with pole subtraction graph by graph away from the collinear limit. Appendix~\ref{app:modified-bg} proves instead, at the level of the complete hard-side current, that the modified recursion reproduces the strict pole-subtracted hard term.

\section{Four-dimensional subleading collinear limits from BCFW recursion}\label{sec:four-dimensional-collinear}
In four dimensions it is convenient to describe the collinear limit using spinor-helicity variables. Let the parent momentum be
\begin{equation}
P = \lambda_P \tilde\lambda_P, \qquad P^2=0
\end{equation}
and we introduce a reference null momentum
\begin{equation}
r = \lambda_r \tilde\lambda_r,
\qquad r^2 = 0 ,
\qquad r\!\cdot P \neq 0 .
\end{equation}
For $0\leq x\leq 1$, it is useful to introduce an angle $\theta$ and the shorthand notation
\begin{equation}
 c:=\cos\theta=\sqrt{x},
 \qquad
 s:=\sin\theta=\sqrt{1-x},
 \qquad c^2+s^2=1.
\label{eq:cs-parametrization}
\end{equation}
The collinear limit of legs $n-1$ and $n$ is then parametrized as
\begin{equation}\label{eq:4d-collinear-parametrization}
\begin{aligned}
\lambda_{n-1} &= c\,\lambda_P
- \epsilon s\,\lambda_r ,\\
\lambda_n &= s\,\lambda_P
+ \epsilon c\,\lambda_r ,
\end{aligned}
\qquad
\begin{aligned}
\tilde\lambda_{n-1} &= c\,\tilde\lambda_P
- \epsilon s\,\tilde\lambda_r ,\\
\tilde\lambda_n &= s\,\tilde\lambda_P
+ \epsilon c\,\tilde\lambda_r .
\end{aligned}
\end{equation}
Equivalently, each chiral sector takes the form
\begin{equation}
 \begin{pmatrix}\lambda_{n-1}\\ \lambda_n\end{pmatrix}
 =
 \begin{pmatrix}c&-s\\ s&c\end{pmatrix}
 \begin{pmatrix}\lambda_P\\ \epsilon\lambda_r\end{pmatrix},
 \qquad
 \begin{pmatrix}\tilde\lambda_{n-1}\\ \tilde\lambda_n\end{pmatrix}
 =
 \begin{pmatrix}c&-s\\ s&c\end{pmatrix}
 \begin{pmatrix}\tilde\lambda_P\\ \epsilon\tilde\lambda_r\end{pmatrix}.
\label{eq:spinor-plane-rotation}
\end{equation}
Thus the parametrization contains an $SO(2)$ rotation in the auxiliary two-dimensional space of parent and rescaled reference-spinor data. The corresponding momenta are
\begin{align}
p_{n-1}
&=xP-\epsilon\sqrt{x(1-x)}
 \bigl(\lambda_P\tilde\lambda_r+\lambda_r\tilde\lambda_P\bigr)
 +\epsilon^2(1-x)r,\nonumber\\
p_n
&=(1-x)P+\epsilon\sqrt{x(1-x)}
 \bigl(\lambda_P\tilde\lambda_r+\lambda_r\tilde\lambda_P\bigr)
 +\epsilon^2x r .
\label{eq:4d-momentum-expansion}
\end{align}
They satisfy
\begin{equation}
p_{n-1}^2 = p_n^2 = 0,
\qquad
(p_{n-1}+p_n)^2 = 2 \epsilon^2 \,P\!\cdot r \,,
\end{equation}
where we have adopted the convention $s_{i,j}=(p_i+p_j)^2=\ab{i}{j}\sqb{i}{j}$.

The four-dimensional spinor-helicity parametrization is obtained from the $D$-dimensional decomposition by restricting to a four-dimensional subspace spanned by $\{P^\mu, K^\mu, e_1^\mu, e_2^\mu\}$, where $e_{1,2}^\mu$ form an orthonormal basis of the transverse space. In four dimensions the transverse momentum can be written as
\begin{equation} \label{eq: 4d kperp}
k_\perp \;\longleftrightarrow\; \sqrt{x(1-x)} (\lambda_P \tilde\lambda_r + \lambda_r \tilde\lambda_P) ,
\end{equation}
and the null reference vector $K$ may be chosen in terms of $r$ as\footnote{
The normalization of the auxiliary null vector \(K\) is not important in
the \(D\)-dimensional parametrization.  Indeed, one may equivalently choose \(K^\mu=a\,r^\mu\) for any nonzero constant \(a\), since \(K^\mu/(P\cdot K)=r^\mu/(P\cdot r)\).  
}
\begin{equation} \label{eq: 4d K}
K \;\longleftrightarrow\; 
\lambda_r \tilde\lambda_r .
\end{equation}
With this normalization, $k_\perp^2=-2x(1-x)P\cdot K$.  Substitution into Eq.~\eqref{eq: collinear momenta} therefore reproduces the quadratic terms $+\epsilon^2(1-x)r$ and $+\epsilon^2x r$ in Eq.~\eqref{eq:4d-momentum-expansion}. Therefore the single reference spinor $r$ in four dimensions encodes both the lightlike direction $K^\mu$ and a choice of transverse two-plane.

The general-dimensional expressions reduce to four dimensions through Eqs.~\eqref{eq: 4d kperp} and~\eqref{eq: 4d K}. Spinor-helicity variables, however, make the helicity dependence and the analytic structure considerably more transparent. We therefore develop a direct four-dimensional construction of the strict subleading term.  The explicit reduction of the covariant equal- and mixed-polarization splitting tensors at the leading collinear limit is collected in Appendix~\ref{app:four-dimensional-reduction}. In this section, we focus on the case in which the two collinear legs have the same helicity. The distinct-helicity cases are discussed in Section~\ref{sec:distinct-polarizations}.

A useful benchmark is the Parke--Taylor formula for MHV amplitudes~\cite{Parke:1986gb}:
\begin{equation}
A(1^+2^+\ldots i^-\ldots j^-\ldots n^+)= \frac{\langle ij\rangle^4}{\langle 12\rangle\langle 23\rangle \ldots \langle n1\rangle}\,.
\end{equation}
For $i,j \neq n-1,n$, its strict subleading collinear term is\footnote{If instead the two negative-helicity legs are precisely the collinear pair, the amplitude $A(1^+,\ldots,(n-2)^+,(n-1)^-,n^-)$ has a vanishing strict $O(\epsilon^0)$ collinear coefficient. The same statement holds for its parity-conjugate $\overline{\mathrm{MHV}}$ amplitude with two positive-helicity collinear legs.}
\begin{equation}
\begin{aligned}
&A(1^+2^+\ldots i^-\ldots j^-\ldots (n-1)^+n^+)\Big|^{(0)}_{\rm coll.}\\
=&\Big(\frac{1}{x}\frac{\ab{r}{n-2}}
{\ab{r}{P}\ab{n-2}{P}}-\frac{1}{1-x}\frac{\ab{r}{1}}
{\ab{1}{P}\ab{r}{P}}\Big)\, A(1^+2^+\ldots i^-\ldots j^- \ldots P^+)\,.
\end{aligned}
\end{equation}
At the symmetric split $x=1/2$, Schouten's identity eliminates the reference-spinor dependence, as expected from the general-dimensional expression~\eqref{eq: subleading}, yielding
\begin{equation} \label{eq: PT collinear}
\begin{aligned}
&A(1^+2^+\ldots i^-\ldots j^-\ldots (n-1)^+n^+)\Big|^{(0)}_{\rm coll.}\\
\xrightarrow[]{x=1/2}&\, -\frac{2\ab{1}{n-2}}{\ab{n-2}{P}\ab{P}{1}}\ A(1^+2^+\ldots i^-\ldots j^- \ldots P^+)\,,
\end{aligned}
\end{equation}
The prefactor is proportional to the universal positive-helicity ``soft'' factor for the parent leg $P$ in the lower-point amplitude.

This MHV result supplies the terminal rule for the recursive construction below.  Most displayed examples use $x=1/2$ to expose their simplest form; the recursion itself and its MHV terminal data remain valid at generic $x$.

At five points only the MHV and $\overline{\rm MHV}$ sectors occur, therefore it is easy to show the subleading collinear limit at $x=1/2$ factorizes into a four-point amplitude multiply by a ``soft'' factor,
\begin{equation}
    A(1,2,3,4,5)\Big|^{(0)}_{\rm coll.}= 2(\frac{\ve_P \cdot p_1}{s_{1,P}}-\frac{\ve_P \cdot p_3}{s_{3,P}}) A(1,2,3,P).
\end{equation}

\subsection{Subleading collinear limit from BCFW recursion}
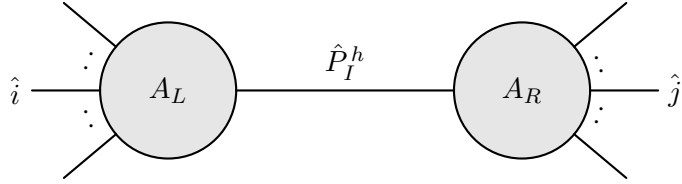
\begin{figure}[h]
\centering
\begin{tikzpicture}[baseline=-0.5ex,scale=0.9,thick,line cap=round,line join=round]
\def\rL{1.0}
\def\rR{1.0}

\draw[fill=gray!20] (0,0) circle (\rL);
\draw[fill=gray!20] (5.2,0) circle (\rR);

\node at (0,0) {$A_L$};
\node at (5.2,0) {$A_R$};

\draw (\rL,0) -- (5.2-\rR,0);
\node at (2.6,0.42) {$\hat P_I^{\,h}$};


\draw (140:\rL) -- (140:2);

\draw (180:\rL) -- (180:2);
\node at (180:2.25) {$\hat{i}$};

\draw (220:\rL) -- (220:2);

\node at (155:1.25) {$\cdot$};
\node at (165:1.25) {$\cdot$};

\node at (195:1.25) {$\cdot$};
\node at (205:1.25) {$\cdot$};


\draw ($(5.2,0)+(40:\rR)$) -- ++(40:1.0);

\draw ($(5.2,0)+(0:\rR)$) -- ++(0:1.0);
\node at (7.45,0) {$\hat{j}$};

\draw ($(5.2,0)+(-40:\rR)$) -- ++(-40:1.0);

\node at ($(5.3,0)+(15:1.1)$) {$\cdot$};
\node at ($(5.3,0)+(25:1.1)$) {$\cdot$};

\node at ($(5.3,0)+(-15:1.1)$) {$\cdot$};
\node at ($(5.3,0)+(-25:1.1)$) {$\cdot$};

\end{tikzpicture}
\caption{A BCFW term obtained by gluing two lower-point on-shell amplitudes
across the state $\hat P_I^{,h}$.}
\label{fig:bcfw_bridge}
\end{figure}

The Britto--Cachazo--Feng--Witten (BCFW) recursion~\cite{Britto:2004ap,Britto:2005fq} provides compact representations for arbitrary helicity sectors.  We use the $[i,j\rangle$ deformation
\begin{equation}
|\hat i]=|i]+z|j],
\qquad
|\hat j\rangle=|j\rangle-z|i\rangle,
\qquad
|\hat i\rangle=|i\rangle,
\qquad
|\hat j]=|j].
\end{equation}
The shifted momenta remain massless and preserve total momentum. Provided that $A_n(z)$ vanishes as $z\to\infty$, Cauchy's theorem reconstructs $A_n(0)$ from the finite poles at which an internal momentum $\hat P_I(z)$ becomes null. In Yang--Mills theory, $[+,-\rangle$-type shifts are excluded because the corresponding amplitude does not vanish at large $z$; see Refs.~\cite{Arkani-Hamed:2008bsc,Cohen:2010mi} for discussions of the large-$z$ behavior.  The recursion reads
\begin{equation}
A_n
=
\sum_{I,h}
A_L\bigl(z_I;\hat P_I^{h}\bigr)\,
\frac{1}{P_I^2}\,
A_R\bigl(z_I;-\hat P_I^{-h}\bigr),
\label{eq:BCFWrecursion}
\end{equation}
where the sum runs over compatible ordered channels and intermediate helicities, and $\hat P_I(z_I)^2=0$.  Figure~\ref{fig:bcfw_bridge} depicts a typical term.

We choose the initially shifted legs away from the collinear pair; lower recursive calls may choose their own valid shifts.  The BCFW terms then fall into two classes:
\begin{enumerate}
    \item If $n-1$ and $n$ lie in different subamplitudes, the BCFW term has
no adjacent collinear pole.  Its strict coefficient is therefore obtained by evaluating that term directly on the collinear surface.  At $x=1/2$ this means
    \begin{equation}
\begin{aligned} \label{eq: x1/2 lambda}
\lambda_{n-1} =
\lambda_n = \frac{\sqrt{2}}{2}\,\lambda_P,
\end{aligned}
\quad
\begin{aligned}
\tilde\lambda_{n-1} =
\tilde\lambda_n = \frac{\sqrt{2}}{2}\,\tilde\lambda_P,
\end{aligned}
\end{equation}
The corresponding diagrammatic replacement is
\begin{equation} \label{eq: 4d case 1}
\left[
\begin{tikzpicture}[baseline=-0.5ex,scale=0.62,thick,line cap=round,line join=round]
\def\rL{0.8}
\def\rR{0.8}

\draw[fill=gray!20] (0,0) circle (\rL);
\draw[fill=gray!20] (3.3,0) circle (\rR);
\node at (0,0) {$A_L$};
\node at (3.3,0) {$A_R$};

\draw (\rL,0) -- (3.3-\rR,0);
\node at (1.65,0.5) {$\hat P_I^{\,h}$};

\draw (140:\rL) -- (140:1.48);
\draw (180:\rL) -- (180:1.48);
\node at (180:1.72) {$\hat{i}$};

\node at (150:0.98) {$\cdot$};
\node at (165:1.00) {$\cdot$};
\node at (195:1.00) {$\cdot$};
\node at (210:0.98) {$\cdot$};

\draw (225:\rL) -- (225:1.58);
\node at (226:1.88) {$n$};

\draw ($(3.3,0)+(40:\rR)$) -- ++(40:0.88);
\draw ($(3.3,0)+(0:\rR)$) -- ++(0:0.88);
\node at (5.2,0) {$\hat{j}$};

\draw ($(3.3,0)+(-40:\rR)$) -- ++(-40:0.88);
\node at ($(3.3,0)+(-40:2)$) {$n\!-\!1$};

\node at ($(3.3,0)+(13.3:1)$) {$\cdot$};
\node at ($(3.3,0)+(26.6:1)$) {$\cdot$};

\node at ($(3.3,0)+(-13.3:1)$) {$\cdot$};
\node at ($(3.3,0)+(-26.6:1)$) {$\cdot$};
\end{tikzpicture}
\right]^{\!(0)}_{\!\mathrm{coll.}}
\;\xrightarrow{x=\frac12}\;
\left[
\begin{tikzpicture}[baseline=-0.5ex,scale=0.62,thick,line cap=round,line join=round]
\def\rL{0.8}
\def\rR{0.8}

\draw[fill=gray!20] (0,0) circle (\rL);
\draw[fill=gray!20] (3.3,0) circle (\rR);
\node at (0,0) {$A_L$};
\node at (3.3,0) {$A_R$};

\draw (\rL,0) -- (3.3-\rR,0);
\node at (1.65,0.5) {$\hat P_I^{\,h}$};

\draw (140:\rL) -- (140:1.48);
\draw (180:\rL) -- (180:1.48);
\node at (180:1.72) {$\hat{i}$};

\node at (150:0.98) {$\cdot$};
\node at (165:1.00) {$\cdot$};
\node at (195:1.00) {$\cdot$};
\node at (210:0.98) {$\cdot$};

\draw (225:\rL) -- (225:1.58);
\node at (226:1.88) {$n$};

\draw ($(3.3,0)+(40:\rR)$) -- ++(40:0.88);
\draw ($(3.3,0)+(0:\rR)$) -- ++(0:0.88);
\node at (5.2,0) {$\hat{j}$};

\draw ($(3.3,0)+(-40:\rR)$) -- ++(-40:0.88);
\node at ($(3.3,0)+(-40:2)$) {$n\!-\!1$};

\node at ($(3.3,0)+(13.3:1)$) {$\cdot$};
\node at ($(3.3,0)+(26.6:1)$) {$\cdot$};

\node at ($(3.3,0)+(-13.3:1)$) {$\cdot$};
\node at ($(3.3,0)+(-26.6:1)$) {$\cdot$};
\end{tikzpicture}
\right]_{\!\!\begin{array}{c}
\scriptstyle \lambda_{n-1}=\lambda_n=\frac{\sqrt2}{2}\lambda_P \\[1pt]
\scriptstyle \tilde\lambda_{n-1}=\tilde\lambda_n=\frac{\sqrt2}{2}\tilde\lambda_P
\end{array}}
\end{equation}

\item If both collinear legs lie in the same subamplitude, say $A_L$, the
strict collinear operation is applied recursively to $A_L$.  Subsequent recursion either separates the pair, reducing to the first case, or reaches an MHV or $\overline{\rm MHV}$ terminal amplitude, where the closed formula above applies.  Diagrammatically,
\begin{equation}
\left[
\begin{tikzpicture}[baseline=-0.5ex,scale=0.62,thick,line cap=round,line join=round]
\def\rL{0.82}
\def\rR{0.82}

\draw[fill=gray!20] (0,0) circle (\rL);
\draw[fill=gray!20] (3.3,0) circle (\rR);

\node at (0,0) {$A_L$};
\node at (3.3,0) {$A_R$};

\draw (\rL,0) -- ({3.3-\rR},0);
\node at (1.8,0.5) {$\hat P_I^{\,h}$};


\draw (90:\rL) -- (90:1.75);
\draw (270:\rL) -- (270:1.75);

\draw (140:\rL) -- (140:1.75);
\node at (140:2.0) {$\hat{i}$};

\draw (180:\rL) -- (180:1.75);
\node at (180:2.0) {$n$};

\draw (200:\rL) -- (200:1.75);
\node at (200:2.5) {$n\!-\!1$};

\node at (105:1) {$\cdot$};
\node at (120:1) {$\cdot$};
\node at (155:1) {$\cdot$};
\node at (168:1) {$\cdot$};
\node at (222:1) {$\cdot$};
\node at (235:1) {$\cdot$};
\node at (250:1) {$\cdot$};


\draw ($(3.3,0)+(40:\rR)$) -- ++(40:1.0);

\draw ($(3.3,0)+(0:\rR)$) -- ++(0:1.0);
\node at (5.4,0) {$\hat{j}$};

\draw ($(3.3,0)+(-40:\rR)$) -- ++(-40:1.0);

\node at ($(3.3,0)+(15:0.97)$) {$\cdot$};
\node at ($(3.3,0)+(28:0.97)$) {$\cdot$};
\node at ($(3.3,0)+(-15:0.97)$) {$\cdot$};
\node at ($(3.3,0)+(-28:0.97)$) {$\cdot$};

\end{tikzpicture}\right]^{\!(0)}_{\!\mathrm{coll.}}=
\begin{tikzpicture}[baseline=-0.5ex,
    scale=0.62, thick, line cap=round, line join=round]

\def\rL{0.82}
\def\rR{0.82}
\def\leg{0.93} 

\coordinate (CL) at (0,0);
\draw[fill=gray!20] (CL) circle (\rL);
\node at (CL) {$A_L$};

\draw ($(CL)+(90:\rL)$)  -- ++(90:\leg);
\draw ($(CL)+(270:\rL)$) -- ++(270:\leg);

\draw ($(CL)+(140:\rL)$) -- ++(140:\leg);
\node at ($(CL)+(140:2.0)$) {$\hat{i}$};

\draw ($(CL)+(180:\rL)$) -- ++(180:\leg);
\node at ($(CL)+(180:2.0)$) {$n$};

\draw ($(CL)+(205:\rL)$) -- ++(205:\leg);
\node at ($(CL)+(210:2.25)$) {$n\!-\!1$};

\node at ($(CL)+(105:1)$) {$\cdot$};
\node at ($(CL)+(120:1)$) {$\cdot$};
\node at ($(CL)+(155:1)$) {$\cdot$};
\node at ($(CL)+(168:1)$) {$\cdot$};
\node at ($(CL)+(222:1)$) {$\cdot$};
\node at ($(CL)+(235:1)$) {$\cdot$};
\node at ($(CL)+(250:1)$) {$\cdot$};

\coordinate (CR) at (4.8,0);
\draw[fill=gray!20] (CR) circle (\rR);
\node at (CR) {$A_R$};

\draw ($(CL)+(\rL,0)$) -- ($(CR)+(-\rR,0)$);
\node at (2.4,0.5) {$\hat P_I^{\,h}$};

\draw ($(CR)+(40:\rR)$)  -- ++(40:\leg);
\draw ($(CR)+(0:\rR)$)   -- ++(0:\leg);
\node at ($(CR)+(0:2.1)$) {$\hat{j}$};
\draw ($(CR)+(-40:\rR)$) -- ++(-40:\leg);

\node at ($(CR)+(15:0.97)$) {$\cdot$};
\node at ($(CR)+(28:0.97)$) {$\cdot$};
\node at ($(CR)+(-15:0.97)$) {$\cdot$};
\node at ($(CR)+(-28:0.97)$) {$\cdot$};

\draw (-2.2, 2.0) -- (-2.6, 2.0) -- (-2.6,-2.0) -- (-2.2,-2.0);
\draw ( 1.3, 2.0) -- ( 1.7, 2.0) -- ( 1.7,-2.0) -- ( 1.3,-2.0);

\node[anchor=west] at (1.75,  1.95) {$^{(0)}$};
\node[anchor=west] at (1.75, -1.95) {$\mathrm{coll.}$};

\end{tikzpicture}
\end{equation}

The recursion terminates because every branch either separates the two legs or reaches an MHV or $\overline{\rm MHV}$ amplitude.  In particular, no Laurent expansion of a fully assembled $n$-point expression is required.

\end{enumerate}

\subsection{Examples}
\subsubsection{An NMHV split-helicity amplitude}
We first consider the NMHV amplitude $A(1^-2^-3^-4^+\ldots n^+)$ under a $[1,2\rangle$ shift.  Its BCFW representation is
\begin{equation} \label{eq: NMHV diagrams}
\begin{aligned}
 &A(1^-2^-3^-4^+\ldots n^+)\\
 =&  \begin{tikzpicture}[baseline=-0.5ex,scale=0.62,thick,line cap=round,line join=round]
\def\rL{0.8}
\def\rR{0.8}

\draw[fill=gray!20] (0,0) circle (\rL);
\draw[fill=gray!20] (3.3,0) circle (\rR);
\node at (0,0) {$A_L$};
\node at (3.3,0) {$A_R$};

\draw (\rL,0) -- (3.3-\rR,0);

\draw (140:\rL) -- (140:1.48);
\node at (135:1.72) {$\hat{1}^-$};


\draw (225:\rL) -- (225:1.58);
\node at (226:1.88) {$n^+$};

\draw ($(3.3,0)+(40:\rR)$) -- ++(40:0.88);
\node at ($(3.3,0)+(40:2.2)$) {$\hat{2}^-$};

\draw ($(3.3,0)+(-40:\rR)$) -- ++(-40:0.88);
\node at ($(3.3,0)+(-40:2)$) {$(n\!-\!1)^+$};

\node at ($(0,0)+(15:1.2)$) {$-$};
\node at ($(3.3,0)+(165:1.2)$) {$+$};
\draw ($(3.3,0)+(20:\rR)$) -- ++(20:1.0);
\node at ($(3.3,0)+(20:2.4)$) {$3^-$};
\node at ($(3.3,0)+(0:1)$) {$\cdot$};
\node at ($(3.3,0)+(-20:1)$) {$\cdot$};
\end{tikzpicture}
+ 
\sum_{i=5}^{n-1} \begin{tikzpicture}[baseline=-0.5ex,scale=0.62,thick,line cap=round,line join=round]
\def\rL{0.82}
\def\rR{0.82}

\draw[fill=gray!20] (0,0) circle (\rL);
\draw[fill=gray!20] (3.3,0) circle (\rR);

\node at (0,0) {$A_L$};
\node at (3.3,0) {$A_R$};

\draw (\rL,0) -- ({3.3-\rR},0);



\draw (140:\rL) -- (140:1.75);
\node at (140:2.0) {$\hat{1}^-$};

\draw (180:\rL) -- (180:1.75);
\node at (180:2.3) {$n^+$};

\draw (200:\rL) -- (200:1.75);
\node at (200:2.9) {$(n-1)^+$};
\node at (240:2.2) {$i^+$};
\draw (240:\rL) -- (240:1.75);

\node at ($(0,0)+(15:1.3)$) {$-$};
\node at ($(3.3,0)+(165:1.2)$) {$+$};

\node at (155:1) {$\cdot$};
\node at (168:1) {$\cdot$};
\node at (215:1) {$\cdot$};
\node at (227:1) {$\cdot$};


\draw ($(3.3,0)+(40:\rR)$) -- ++(40:1.0);

\node at ($(3.3,0)+(40:2.4)$) {$\hat{2}^-$};
\node at ($(3.3,0)+(-40:2.4)$) {$(i\!-\!1)^+$};

\draw ($(3.3,0)+(-40:\rR)$) -- ++(-40:1.0);
\draw ($(3.3,0)+(20:\rR)$) -- ++(20:1.0);
\node at ($(3.3,0)+(20:2.4)$) {$3^-$};

\node at ($(3.3,0)+(0:1)$) {$\cdot$};
\node at ($(3.3,0)+(-20:1)$) {$\cdot$};

\end{tikzpicture}
\end{aligned}
\end{equation}
The first term separates the collinear legs, whereas every term in the sum places them in the same subamplitude.  The separated contribution is
\begin{equation}
    \begin{tikzpicture}[baseline=-0.5ex,scale=0.62,thick,line cap=round,line join=round]
\def\rL{0.8}
\def\rR{0.8}

\draw[fill=gray!20] (0,0) circle (\rL);
\draw[fill=gray!20] (3.3,0) circle (\rR);
\node at (0,0) {$A_L$};
\node at (3.3,0) {$A_R$};

\draw (\rL,0) -- (3.3-\rR,0);

\draw (140:\rL) -- (140:1.48);
\node at (135:1.72) {$\hat{1}^-$};


\draw (225:\rL) -- (225:1.58);
\node at (226:1.88) {$n^+$};

\draw ($(3.3,0)+(40:\rR)$) -- ++(40:0.88);
\node at ($(3.3,0)+(40:2.2)$) {$\hat{2}^-$};

\draw ($(3.3,0)+(-40:\rR)$) -- ++(-40:0.88);
\node at ($(3.3,0)+(-40:2)$) {$(n\!-\!1)^+$};

\node at ($(0,0)+(15:1.2)$) {$-$};
\node at ($(3.3,0)+(165:1.2)$) {$+$};
\draw ($(3.3,0)+(20:\rR)$) -- ++(20:1.0);
\node at ($(3.3,0)+(20:2.4)$) {$3^-$};
\node at ($(3.3,0)+(0:1)$) {$\cdot$};
\node at ($(3.3,0)+(-20:1)$) {$\cdot$};
\end{tikzpicture}= \frac{\langle 3| p_1+p_2|n]^3}{\sqb{n}{1}\sqb{1}{2}\ab{3}{4}\ab{4}{5}\ldots\ab{n\!-\!2}{n\!-\!1} \langle n\!-\!1| p_1+p_{n-1}|2]s_{1,2,n}}
\end{equation}
Applying the replacement~\eqref{eq: x1/2 lambda} to this expression gives
\begin{equation}
    \frac{\langle 3| p_1+p_2|P]^3}{\sqb{P}{1}\sqb{1}{2}\ab{3}{4}\ab{4}{5}\ldots\ab{n\!-\!2}{P} \langle P| p_1|2]\ \frac{1}{2}(s_{1,2}+s_{1,2,P})}
\end{equation}
This expression is proportional to the corresponding lower-point diagram $(1^-2^-3^-4^+\ldots (n\!-\!2)^+P^+)$, which we denote by
\begin{equation} \label{eq: NMHV example term1}
\begin{aligned}
&\left[    \begin{tikzpicture}[baseline=-0.5ex,scale=0.62,thick,line cap=round,line join=round]
\def\rL{0.8}
\def\rR{0.8}

\draw[fill=gray!20] (0,0) circle (\rL);
\draw[fill=gray!20] (3.3,0) circle (\rR);
\node at (0,0) {$A_L$};
\node at (3.3,0) {$A_R$};

\draw (\rL,0) -- (3.3-\rR,0);

\draw (140:\rL) -- (140:1.48);
\node at (135:1.72) {$\hat{1}^-$};


\draw (225:\rL) -- (225:1.58);
\node at (226:1.88) {$n^+$};

\draw ($(3.3,0)+(40:\rR)$) -- ++(40:0.88);
\node at ($(3.3,0)+(40:2.2)$) {$\hat{2}^-$};

\draw ($(3.3,0)+(-40:\rR)$) -- ++(-40:0.88);
\node at ($(3.3,0)+(-40:2)$) {$(n\!-\!1)^+$};

\node at ($(0,0)+(15:1.2)$) {$-$};
\node at ($(3.3,0)+(165:1.2)$) {$+$};
\draw ($(3.3,0)+(20:\rR)$) -- ++(20:1.0);
\node at ($(3.3,0)+(20:2.4)$) {$3^-$};
\node at ($(3.3,0)+(0:1)$) {$\cdot$};
\node at ($(3.3,0)+(-20:1)$) {$\cdot$};
\end{tikzpicture}
\right]_{\!\!\begin{array}{c}
\scriptstyle \lambda_{n-1}=\lambda_n=\frac{\sqrt2}{2}\lambda_P \\[1pt]
\scriptstyle \tilde\lambda_{n-1}=\tilde\lambda_n=\frac{\sqrt2}{2}\tilde\lambda_P
\end{array}} \\
=& \frac{2\, s_{1,2,P}\langle n\!-\!2|p_1+P |2]}{\sqb{1}{2} \ab{P}{1} \ab{n\!-\!2}{P} (s_{1,2}+s_{1,2,P})}\ \times \begin{tikzpicture}[baseline=-0.5ex,scale=0.62,thick,line cap=round,line join=round]
\def\rL{0.8}
\def\rR{0.8}

\draw[fill=gray!20] (0,0) circle (\rL);
\draw[fill=gray!20] (3.3,0) circle (\rR);
\node at (0,0) {$A_L$};
\node at (3.3,0) {$A_R$};

\draw (\rL,0) -- (3.3-\rR,0);

\draw (140:\rL) -- (140:1.48);
\node at (135:1.72) {$\hat{1}^-$};


\draw (225:\rL) -- (225:1.58);
\node at (226:2) {$P^+$};

\draw ($(3.3,0)+(40:\rR)$) -- ++(40:0.88);
\node at ($(3.3,0)+(40:2.2)$) {$\hat{2}^-$};

\draw ($(3.3,0)+(-40:\rR)$) -- ++(-40:0.88);
\node at ($(3.3,0)+(-40:2)$) {$(n\!-\!2)^+$};

\node at ($(0,0)+(15:1.2)$) {$-$};
\node at ($(3.3,0)+(165:1.2)$) {$+$};
\draw ($(3.3,0)+(20:\rR)$) -- ++(20:1.0);
\node at ($(3.3,0)+(20:2.4)$) {$3^-$};
\node at ($(3.3,0)+(0:1)$) {$\cdot$};
\node at ($(3.3,0)+(-20:1)$) {$\cdot$};
\end{tikzpicture}
\end{aligned}\end{equation}

Every left subamplitude in the remaining sum is MHV.  Applying the terminal rule \eqref{eq: PT collinear} gives

\begin{equation}
\begin{aligned}
&\left[    \begin{tikzpicture}[baseline=-0.5ex,scale=0.62,thick,line cap=round,line join=round]
\def\rL{0.82}
\def\rR{0.82}

\draw[fill=gray!20] (0,0) circle (\rL);
\draw[fill=gray!20] (3.3,0) circle (\rR);

\node at (0,0) {$A_L$};
\node at (3.3,0) {$A_R$};

\draw (\rL,0) -- ({3.3-\rR},0);


\draw (140:\rL) -- (140:1.75);
\node at (140:2.0) {$\hat{1}^-$};

\draw (160:\rL) -- (160:1.75);
\node at (160:2.3) {$n^+$};

\draw (180:\rL) -- (180:1.75);
\node at (180:2.9) {$(n-1)^+$};
\node at (240:2.2) {$i^+$};
\draw (240:\rL) -- (240:1.75);

\node at ($(0,0)+(15:1.3)$) {$-$};
\node at ($(3.3,0)+(165:1.2)$) {$+$};

\node at (205:1) {$\cdot$};
\node at (225:1) {$\cdot$};


\draw ($(3.3,0)+(40:\rR)$) -- ++(40:1.0);

\node at ($(3.3,0)+(40:2.4)$) {$\hat{2}^-$};
\node at ($(3.3,0)+(-40:2.4)$) {$(i\!-\!1)^+$};

\draw ($(3.3,0)+(-40:\rR)$) -- ++(-40:1.0);
\draw ($(3.3,0)+(20:\rR)$) -- ++(20:1.0);
\node at ($(3.3,0)+(20:2.4)$) {$3^-$};

\node at ($(3.3,0)+(0:1)$) {$\cdot$};
\node at ($(3.3,0)+(-20:1)$) {$\cdot$};

\end{tikzpicture}
\right]^{\!(0)}_{\!\mathrm{coll.}} \xrightarrow[]{x=\frac{1}{2}}\\
& \frac{2\ab{1}{n\!-\!2}}{\ab{1}{P}\ab{n\!-\!2}{P}}\ \times 
\begin{tikzpicture}[baseline=-0.5ex,scale=0.62,thick,line cap=round,line join=round]
\def\rL{0.82}
\def\rR{0.82}

\draw[fill=gray!20] (0,0) circle (\rL);
\draw[fill=gray!20] (3.3,0) circle (\rR);

\node at (0,0) {$A_L$};
\node at (3.3,0) {$A_R$};

\draw (\rL,0) -- ({3.3-\rR},0);


\draw (140:\rL) -- (140:1.75);
\node at (140:2.0) {$\hat{1}^-$};


\draw (170:\rL) -- (170:1.75);
\node at (170:2.3) {$P^+$};

\draw (230:\rL) -- (230:1.75);
\node at (230:2.2) {$i^+$};

\node at ($(0,0)+(15:1.3)$) {$-$};
\node at ($(3.3,0)+(165:1.2)$) {$+$};

\node at (190:1) {$\cdot$};
\node at (210:1) {$\cdot$};


\draw ($(3.3,0)+(40:\rR)$) -- ++(40:1.0);

\node at ($(3.3,0)+(40:2.4)$) {$\hat{2}^-$};
\node at ($(3.3,0)+(-40:2.4)$) {$(i\!-\!1)^+$};

\draw ($(3.3,0)+(-40:\rR)$) -- ++(-40:1.0);
\draw ($(3.3,0)+(20:\rR)$) -- ++(20:1.0);
\node at ($(3.3,0)+(20:2.4)$) {$3^-$};

\node at ($(3.3,0)+(0:1)$) {$\cdot$};
\node at ($(3.3,0)+(-20:1)$) {$\cdot$};

\end{tikzpicture}
\end{aligned}
\end{equation}
for $5 \leq i \leq n-2$. The case $i = n-1$ is slightly special, since the prefactor involves the internal leg, and we obtain:

\begin{equation} \label{eq: NMHV example term2 special}
\begin{aligned}
&\left[
 \begin{tikzpicture}[baseline=-0.5ex,scale=0.62,thick,line cap=round,line join=round]
\def\rL{0.8}
\def\rR{0.8}

\draw[fill=gray!20] (0,0) circle (\rL);
\draw[fill=gray!20] (3.3,0) circle (\rR);
\node at (0,0) {$A_L$};
\node at (3.3,0) {$A_R$};

\draw (\rL,0) -- (3.3-\rR,0);

\draw (140:\rL) -- (140:1.48);
\node at (135:1.72) {$\hat{1}^-$};


\draw (180:\rL) -- (180:1.58);
\node at (180:2) {$n^+$};

\draw (225:\rL) -- (225:1.58);
\node at (226:2) {$(n\!-\!1)^+$};

\draw ($(3.3,0)+(40:\rR)$) -- ++(40:0.88);
\node at ($(3.3,0)+(40:2.2)$) {$\hat{2}^-$};

\draw ($(3.3,0)+(-40:\rR)$) -- ++(-40:0.88);
\node at ($(3.3,0)+(-40:2)$) {$(n\!-\!1)^+$};

\node at ($(0,0)+(15:1.2)$) {$-$};
\node at ($(3.3,0)+(165:1.2)$) {$+$};
\draw ($(3.3,0)+(20:\rR)$) -- ++(20:1.0);
\node at ($(3.3,0)+(20:2.4)$) {$3^-$};
\node at ($(3.3,0)+(0:1)$) {$\cdot$};
\node at ($(3.3,0)+(-20:1)$) {$\cdot$};
\end{tikzpicture}
\right]^{\!(0)}_{\!\mathrm{coll.}} \xrightarrow[]{x=\frac{1}{2}}\\
& \frac{2\ab{1}{\hat{I}}}{\ab{1}{P}\ab{\hat{I}}{P}}\ \times 
\begin{tikzpicture}[baseline=-0.5ex,scale=0.62,thick,line cap=round,line join=round]
\def\rL{0.82}
\def\rR{0.82}

\draw[fill=gray!20] (0,0) circle (\rL);
\draw[fill=gray!20] (3.3,0) circle (\rR);

\node at (0,0) {$A_L$};
\node at (3.3,0) {$A_R$};

\draw (\rL,0) -- ({3.3-\rR},0);


\draw (140:\rL) -- (140:1.75);
\node at (140:2.0) {$\hat{1}^-$};



\draw (230:\rL) -- (230:1.75);
\node at (230:2.2) {$P^+$};

\node at ($(0,0)+(15:1.3)$) {$-$};
\node at ($(3.3,0)+(165:1.2)$) {$+$};


\draw ($(3.3,0)+(40:\rR)$) -- ++(40:1.0);

\node at ($(3.3,0)+(40:2.4)$) {$\hat{2}^-$};
\node at ($(3.3,0)+(-40:2.4)$) {$(n\!-\!2)^+$};

\draw ($(3.3,0)+(-40:\rR)$) -- ++(-40:1.0);
\draw ($(3.3,0)+(20:\rR)$) -- ++(20:1.0);
\node at ($(3.3,0)+(20:2.4)$) {$3^-$};

\node at ($(3.3,0)+(0:1)$) {$\cdot$};
\node at ($(3.3,0)+(-20:1)$) {$\cdot$};

\end{tikzpicture}
\end{aligned}
\end{equation}
The same lower-point diagram appears in Eq.~\eqref{eq: NMHV example term1}.  Multiplying the prefactor by $\sqb{\hat I}{2}/\sqb{\hat I}{2}$ yields
\begin{equation}
    \frac{2\ab{1}{\hat{I}}\sqb{\hat{I}}{2}}{\ab{1}{P}\ab{\hat{I}}{P}\sqb{\hat{I}}{2}}=\frac{2 \langle\hat{1}|\hat{p}_1+P|2]}{\ab{1}{P}\langle P|\hat{p}_1+P|2]}=\frac{2 \sqb{P}{2}}{\ab{1}{P}\sqb{1}{2}}
\end{equation}
Equations~\eqref{eq: NMHV example term1} and \eqref{eq: NMHV example term2 special} can therefore be combined at fixed lower-point diagram, with prefactor
\begin{equation} \label{eq: combine factor}
    \frac{2\, s_{1,2,P}\langle n\!-\!2|p_1+P |2]}{\sqb{1}{2} \ab{P}{1} \ab{n\!-\!2}{P} (s_{1,2}+s_{1,2,P})}+\frac{2 \sqb{P}{2}}{\ab{1}{P}\sqb{1}{2}}=\frac{2\ab{1}{2}\langle n\!-\!2|p_1+P|2]}{\ab{1}{P}\ab{n\!-\!2}{P}(s_{1,2}+s_{1,2,P})}+\frac{2\ab{1}{n\!-\!2}}{\ab{1}{P}\ab{n\!-\!2}{P}}\,,
\end{equation}
The second form isolates a soft-factor-like contribution that will be useful below.

The strict subleading limit is consequently a sum of lower-point BCFW diagrams multiplied by the corresponding combined prefactors:
\begin{equation}
\begin{aligned}
    &A(1^-2^-3^-4^+\ldots n^+)\Big|^{(0)}_{\rm coll.} \xrightarrow[]{x=\frac12} \\
    & \left(\frac{2\ab{1}{2}\langle n\!-\!2|p_1+P|2]}{\ab{1}{P}\ab{n\!-\!2}{P}(s_{1,2}+s_{1,2,P})}+\frac{2\ab{1}{n\!-\!2}}{\ab{1}{P}\ab{n\!-\!2}{P}}\right)\times \begin{tikzpicture}[baseline=-0.5ex,scale=0.62,thick,line cap=round,line join=round]
\def\rL{0.8}
\def\rR{0.8}

\draw[fill=gray!20] (0,0) circle (\rL);
\draw[fill=gray!20] (3.3,0) circle (\rR);
\node at (0,0) {$A_L$};
\node at (3.3,0) {$A_R$};

\draw (\rL,0) -- (3.3-\rR,0);

\draw (140:\rL) -- (140:1.48);
\node at (135:1.72) {$\hat{1}^-$};


\draw (225:\rL) -- (225:1.58);
\node at (226:2) {$P^+$};

\draw ($(3.3,0)+(40:\rR)$) -- ++(40:0.88);
\node at ($(3.3,0)+(40:2.2)$) {$\hat{2}^-$};

\draw ($(3.3,0)+(-40:\rR)$) -- ++(-40:0.88);
\node at ($(3.3,0)+(-40:2)$) {$(n\!-\!2)^+$};

\node at ($(0,0)+(15:1.2)$) {$-$};
\node at ($(3.3,0)+(165:1.2)$) {$+$};
\draw ($(3.3,0)+(20:\rR)$) -- ++(20:1.0);
\node at ($(3.3,0)+(20:2.4)$) {$3^-$};
\node at ($(3.3,0)+(0:1)$) {$\cdot$};
\node at ($(3.3,0)+(-20:1)$) {$\cdot$};
\end{tikzpicture} \\
&+ \frac{2\ab{1}{n\!-\!2}}{\ab{1}{P}\ab{n\!-\!2}{P}}\ \times  \sum_{i=5}^{n\!-\!2}
\begin{tikzpicture}[baseline=-0.5ex,scale=0.62,thick,line cap=round,line join=round]
\def\rL{0.82}
\def\rR{0.82}

\draw[fill=gray!20] (0,0) circle (\rL);
\draw[fill=gray!20] (3.3,0) circle (\rR);

\node at (0,0) {$A_L$};
\node at (3.3,0) {$A_R$};

\draw (\rL,0) -- ({3.3-\rR},0);


\draw (140:\rL) -- (140:1.48);
\node at (140:1.72) {$\hat{1}^-$};


\draw (170:\rL) -- (170:1.48);
\node at (170:2.0) {$P^+$};

\draw (230:\rL) -- (230:1.7);
\node at (230:2.3) {$i^+$};

\node at ($(0,0)+(15:1.3)$) {$-$};
\node at ($(3.3,0)+(165:1.2)$) {$+$};

\node at (190:1) {$\cdot$};
\node at (210:1) {$\cdot$};


\draw ($(3.3,0)+(40:\rR)$) -- ++(40:1.0);

\node at ($(3.3,0)+(40:2.4)$) {$\hat{2}^-$};
\node at ($(3.3,0)+(-40:2.4)$) {$(i\!-\!1)^+$};

\draw ($(3.3,0)+(-40:\rR)$) -- ++(-40:1.0);
\draw ($(3.3,0)+(20:\rR)$) -- ++(20:1.0);
\node at ($(3.3,0)+(20:2.4)$) {$3^-$};

\node at ($(3.3,0)+(0:1)$) {$\cdot$};
\node at ($(3.3,0)+(-20:1)$) {$\cdot$};

\end{tikzpicture}
\end{aligned}
\end{equation}
The result is nearly proportional to a lower-point amplitude, but the second and third lines carry different prefactors.  It is therefore useful to isolate a soft-factor-like term multiplying the lower-point amplitude and a correction supported on a single BCFW diagram:
\begin{equation} \label{eq: split NMHV}
\begin{aligned}
    &A(1^-2^-3^-4^+\ldots n^+)\Big|^{(0)}_{\rm coll.} \xrightarrow[]{x=\frac12} \frac{2\ab{1}{n\!-\!2}}{\ab{1}{P}\ab{n\!-\!2}{P}} A(1^-2^-3^-4^+\ldots P^+) \\
    &+ \frac{2\ab{1}{2}\langle n\!-\!2|p_1+P|2]}{\ab{1}{P}\ab{n\!-\!2}{P}(s_{1,2}+s_{1,2,P})}\ \times \begin{tikzpicture}[baseline=-0.5ex,scale=0.62,thick,line cap=round,line join=round]
\def\rL{0.8}
\def\rR{0.8}

\draw[fill=gray!20] (0,0) circle (\rL);
\draw[fill=gray!20] (3.3,0) circle (\rR);
\node at (0,0) {$A_L$};
\node at (3.3,0) {$A_R$};

\draw (\rL,0) -- (3.3-\rR,0);

\draw (140:\rL) -- (140:1.48);
\node at (135:1.72) {$\hat{1}^-$};


\draw (225:\rL) -- (225:1.58);
\node at (226:2) {$P^+$};

\draw ($(3.3,0)+(40:\rR)$) -- ++(40:0.88);
\node at ($(3.3,0)+(40:2.2)$) {$\hat{2}^-$};

\draw ($(3.3,0)+(-40:\rR)$) -- ++(-40:0.88);
\node at ($(3.3,0)+(-40:2)$) {$(n\!-\!2)^+$};

\node at ($(0,0)+(15:1.2)$) {$-$};
\node at ($(3.3,0)+(165:1.2)$) {$+$};
\draw ($(3.3,0)+(20:\rR)$) -- ++(20:1.0);
\node at ($(3.3,0)+(20:2.4)$) {$3^-$};
\node at ($(3.3,0)+(0:1)$) {$\cdot$};
\node at ($(3.3,0)+(-20:1)$) {$\cdot$};
\end{tikzpicture}
\end{aligned}
\end{equation}
At $n=6$ only one diagram contributes, and the factorization is exact:
\begin{equation}
\begin{aligned}
A(1^-2^-3^-4^+5^+ 6^+)\Big|^{(0)}_{\rm coll.} \xrightarrow[]{x=\frac12} &
     \left(\frac{2\ab{1}{2}\langle 4|p_1+P|2]}{\ab{1}{P}\ab{4}{P}(s_{1,2}+s_{1,2,P})}+\frac{2\ab{1}{4}}{\ab{1}{P}\ab{4}{P}}\right)  A(1^-2^-3^-4^+P^+) \\
     =&   \left(\frac{2 \ab{1}{4} }{\ab{1}{P}\ab{4}{P}}-\frac{2\ab{1}{2}\ab{3}{4}\sqb{2}{3}}{\ab{1}{P}\ab{4}{P}(s_{1,2}+s_{3,4})}\right)  A(1^-2^-3^-4^+P^+)
\end{aligned}
\end{equation}
In fact, for $n=6$ all NMHV amplitudes with two positive-helicity collinear legs share the same subleading prefactor:
\begin{equation}  \label{eq: 6-pt NMHV general}
\begin{aligned}
A^{\rm NMHV}(1^{h_1}2^{h_2}3^{h_3}4^{h_4}5^+ 6^+)\Big|^{(0)}_{\rm coll.} \xrightarrow[]{x=\frac12} 
        \left(\frac{2 \ab{1}{4} }{\ab{1}{P}\ab{4}{P}}-\frac{2\ab{1}{2}\ab{3}{4}\sqb{2}{3}}{\ab{1}{P}\ab{4}{P}(s_{1,2}+s_{3,4})}\right)  A^{\overline{\rm MHV}}(1^{h_1}2^{h_2}3^{h_3}4^{h_4} P^+)
\end{aligned}
\end{equation}

For generic $x$, with $c=\sqrt{x}$ and $s=\sqrt{1-x}$ defined above, the six-point result becomes
\begin{equation}\label{eq:6pt-NMHV-generic-x}
\begin{aligned}
&A^{\rm NMHV}(1^{h_1}2^{h_2}3^{h_3}4^{h_4}5^+ 6^+)\Big|^{(0)}_{\rm coll.}
=\frac{1}{s^2 \langle 1\,P\rangle \langle 4\,P\rangle \langle r\,P\rangle}\ \Big\{
\frac{s^2 \langle 1\,P\rangle \langle r\,4\rangle-c^2 \langle 4\,P\rangle \langle r\,1\rangle}{c^2}\\
&\quad+\frac{c^2 s_{12}[3\,2]\langle 1\,2\rangle
(\langle 4\,P\rangle\langle r\,3\rangle-\langle 3\,P\rangle\langle r\,4\rangle)}
{s_{34}\left(c^2s_{12}+s^2s_{34}\right)}+\frac{[3\,2]}{[2\,1][P\,1]}
\Big[\frac{(s_{12}+s_{34})[P\,1]}{[4\,3]}\langle r\,P\rangle
+[2\,1]\langle 3\,4\rangle\langle r\,2\rangle \\
&\quad+([2\,1]\langle 2\,3\rangle+[P\,1]\langle 3\,P\rangle)
\langle r\,4\rangle
-([2\,1]\langle 2\,4\rangle+[P\,1]\langle 4\,P\rangle)
\langle r\,3\rangle\Big]\Big\}
A^{\overline{\rm MHV}}(1^{h_1}2^{h_2}3^{h_3}4^{h_4} P^+)\, .
\end{aligned}
\end{equation}

\subsubsection{Split-helicity amplitudes}
The same structure extends to $A(1^-2^-\ldots a^-(a+1)^+\ldots n^+)$, for which an explicit expression is given in~\cite{Britto:2005dg}. Under the $[1,2\rangle$ shift, one BCFW term separates the collinear legs and the remaining $n-5$ terms place them in a common MHV subamplitude, as in Eq.~\eqref{eq: NMHV diagrams}.  Applying respectively Eq.~\eqref{eq: x1/2 lambda} and the MHV terminal rule gives

\begin{equation}
\begin{aligned}
    &A(1^-2^-\ldots a^- (a+1)^+\ldots n^+)\Big|^{(0)}_{\rm coll.} \xrightarrow[]{x=\frac12} \frac{2\ab{1}{n\!-\!2}}{\ab{1}{P}\ab{n\!-\!2}{P}} A(1^-2^-\ldots a^- (a+1)^+\ldots (n-2)^+ P^+) \\
    &+ \left(u_a+\frac{2 \sqb{P}{2}}{\ab{1}{P}\sqb{1}{2}}- \frac{2\ab{1}{n\!-\!2}}{\ab{1}{P}\ab{n\!-\!2}{P}}\right)\ \times \begin{tikzpicture}[baseline=-0.5ex,scale=0.62,thick,line cap=round,line join=round]
\def\rL{0.8}
\def\rR{0.8}

\draw[fill=gray!20] (0,0) circle (\rL);
\draw[fill=gray!20] (3.3,0) circle (\rR);
\node at (0,0) {$A_L$};
\node at (3.3,0) {$A_R$};

\draw (\rL,0) -- (3.3-\rR,0);

\draw (140:\rL) -- (140:1.48);
\node at (135:1.72) {$\hat{1}^-$};


\draw (225:\rL) -- (225:1.58);
\node at (226:2) {$P^+$};

\draw ($(3.3,0)+(40:\rR)$) -- ++(40:0.88);
\node at ($(3.3,0)+(40:2.2)$) {$\hat{2}^-$};

\draw ($(3.3,0)+(-40:\rR)$) -- ++(-40:0.88);
\node at ($(3.3,0)+(-40:2)$) {$(n\!-\!2)^+$};

\node at ($(0,0)+(15:1.2)$) {$-$};
\node at ($(3.3,0)+(165:1.2)$) {$+$};
\draw ($(3.3,0)+(10:\rR)$) -- ++(10:1.0);
\node at ($(3.3,0)+(10:2.4)$) {$a^-$};

\draw ($(3.3,0)+(-10:\rR)$) -- ++(-10:1.0);
\node at ($(3.3,0)+(-10:3)$) {$(a\!+\!1)^+$};
\node at ($(3.3,0)+(30:1)$) {$\cdot$};
\node at ($(3.3,0)+(20:1)$) {$\cdot$};
\node at ($(3.3,0)+(-30:1)$) {$\cdot$};
\node at ($(3.3,0)+(-20:1)$) {$\cdot$};
\end{tikzpicture}
\end{aligned}
\end{equation}
where $u_a$ denotes the ratio
\begin{equation}
    u_a= \left[
\begin{tikzpicture}[baseline=-0.5ex,scale=0.62,thick,line cap=round,line join=round]
\def\rL{0.8}
\def\rR{0.8}

\draw[fill=gray!20] (0,0) circle (\rL);
\draw[fill=gray!20] (3.3,0) circle (\rR);
\node at (0,0) {$A_L$};
\node at (3.3,0) {$A_R$};

\draw (\rL,0) -- (3.3-\rR,0);

\draw (140:\rL) -- (140:1.48);
\node at (135:1.72) {$\hat{1}^-$};


\draw (225:\rL) -- (225:1.58);
\node at (226:2) {$n^+$};

\draw ($(3.3,0)+(40:\rR)$) -- ++(40:0.88);
\node at ($(3.3,0)+(40:2.2)$) {$\hat{2}^-$};

\draw ($(3.3,0)+(-40:\rR)$) -- ++(-40:0.88);
\node at ($(3.3,0)+(-40:2)$) {$(n\!-\!1)^+$};

\node at ($(0,0)+(15:1.2)$) {$-$};
\node at ($(3.3,0)+(165:1.2)$) {$+$};
\draw ($(3.3,0)+(10:\rR)$) -- ++(10:1.0);
\node at ($(3.3,0)+(10:2.4)$) {$a^-$};

\draw ($(3.3,0)+(-10:\rR)$) -- ++(-10:1.0);
\node at ($(3.3,0)+(-10:3)$) {$(a\!+\!1)^+$};
\node at ($(3.3,0)+(30:1)$) {$\cdot$};
\node at ($(3.3,0)+(20:1)$) {$\cdot$};
\node at ($(3.3,0)+(-30:1)$) {$\cdot$};
\node at ($(3.3,0)+(-20:1)$) {$\cdot$};
\end{tikzpicture}\right]_{\!\!\begin{array}{c}
\scriptstyle \lambda_{n-1}=\lambda_n=\frac{\sqrt2}{2}\lambda_P \\[1pt]
\scriptstyle \tilde\lambda_{n-1}=\tilde\lambda_n=\frac{\sqrt2}{2}\tilde\lambda_P
\end{array}}  \Big/
\begin{tikzpicture}[baseline=-0.5ex,scale=0.62,thick,line cap=round,line join=round]
\def\rL{0.8}
\def\rR{0.8}

\draw[fill=gray!20] (0,0) circle (\rL);
\draw[fill=gray!20] (3.3,0) circle (\rR);
\node at (0,0) {$A_L$};
\node at (3.3,0) {$A_R$};

\draw (\rL,0) -- (3.3-\rR,0);

\draw (140:\rL) -- (140:1.48);
\node at (135:1.72) {$\hat{1}^-$};


\draw (225:\rL) -- (225:1.58);
\node at (226:2) {$P^+$};

\draw ($(3.3,0)+(40:\rR)$) -- ++(40:0.88);
\node at ($(3.3,0)+(40:2.2)$) {$\hat{2}^-$};

\draw ($(3.3,0)+(-40:\rR)$) -- ++(-40:0.88);
\node at ($(3.3,0)+(-40:2)$) {$(n\!-\!2)^+$};

\node at ($(0,0)+(15:1.2)$) {$-$};
\node at ($(3.3,0)+(165:1.2)$) {$+$};
\draw ($(3.3,0)+(10:\rR)$) -- ++(10:1.0);
\node at ($(3.3,0)+(10:2.4)$) {$a^-$};

\draw ($(3.3,0)+(-10:\rR)$) -- ++(-10:1.0);
\node at ($(3.3,0)+(-10:3)$) {$(a\!+\!1)^+$};
\node at ($(3.3,0)+(30:1)$) {$\cdot$};
\node at ($(3.3,0)+(20:1)$) {$\cdot$};
\node at ($(3.3,0)+(-30:1)$) {$\cdot$};
\node at ($(3.3,0)+(-20:1)$) {$\cdot$};
\end{tikzpicture}
\end{equation}
\subsubsection{A seven-point NMHV example and its generalization}
As a nonsplit example, consider $A(1^-2^-3^+4^+5^-6^+7^+)$.  The $[1,2\rangle$ shift produces four BCFW terms:
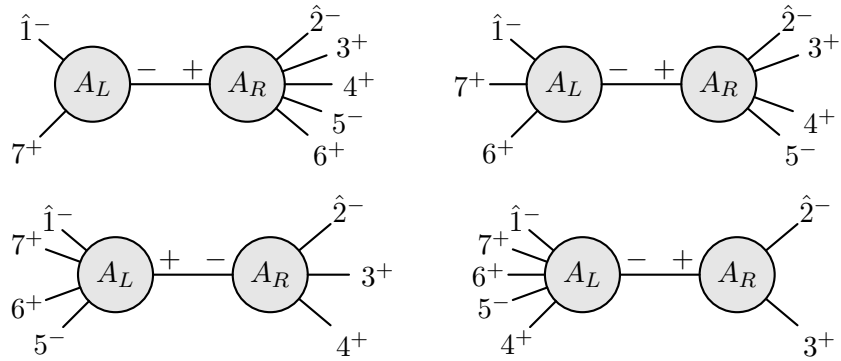
\begin{equation*}
\begin{aligned}
&\begin{tikzpicture}[baseline=-0.5ex,scale=0.62,thick,line cap=round,line join=round]
\def\rL{0.8}
\def\rR{0.8}

\draw[fill=gray!20] (0,0) circle (\rL);
\draw[fill=gray!20] (3.3,0) circle (\rR);
\node at (0,0) {$A_L$};
\node at (3.3,0) {$A_R$};

\draw (\rL,0) -- (3.3-\rR,0);

\draw (140:\rL) -- (140:1.48);
\node at (135:1.72) {$\hat{1}^-$};


\draw (225:\rL) -- (225:1.58);
\node at (226:2) {$7^+$};

\draw ($(3.3,0)+(40:\rR)$) -- ++(40:0.88);
\node at ($(3.3,0)+(40:2.2)$) {$\hat{2}^-$};

\draw ($(3.3,0)+(-40:\rR)$) -- ++(-40:0.88);
\node at ($(3.3,0)+(-40:2.3)$) {$6^+$};

\draw ($(3.3,0)+(-20:\rR)$) -- ++(-20:0.88);
\node at ($(3.3,0)+(-20:2.3)$) {$5^-$};

\node at ($(0,0)+(15:1.2)$) {$-$};
\node at ($(3.3,0)+(165:1.2)$) {$+$};
\draw ($(3.3,0)+(20:\rR)$) -- ++(20:1.0);
\node at ($(3.3,0)+(20:2.4)$) {$3^+$};
\draw ($(3.3,0)+(0:\rR)$) -- ++(0:1.0);
\node at ($(3.3,0)+(0:2.4)$) {$4^+$};
\end{tikzpicture}
\qquad
\begin{tikzpicture}[baseline=-0.5ex,scale=0.62,thick,line cap=round,line join=round]
\def\rL{0.8}
\def\rR{0.8}

\draw[fill=gray!20] (0,0) circle (\rL);
\draw[fill=gray!20] (3.3,0) circle (\rR);
\node at (0,0) {$A_L$};
\node at (3.3,0) {$A_R$};

\draw (\rL,0) -- (3.3-\rR,0);

\draw (140:\rL) -- (140:1.48);
\node at (135:1.72) {$\hat{1}^-$};


\draw (180:\rL) -- (180:1.58);
\node at (180:2) {$7^+$};

\draw (225:\rL) -- (225:1.58);
\node at (226:2) {$6^+$};

\draw ($(3.3,0)+(40:\rR)$) -- ++(40:0.88);
\node at ($(3.3,0)+(40:2.2)$) {$\hat{2}^-$};

\draw ($(3.3,0)+(-40:\rR)$) -- ++(-40:0.88);
\node at ($(3.3,0)+(-40:2.3)$) {$5^-$};

\draw ($(3.3,0)+(-20:\rR)$) -- ++(-20:0.88);
\node at ($(3.3,0)+(-20:2.3)$) {$4^+$};

\node at ($(0,0)+(15:1.2)$) {$-$};
\node at ($(3.3,0)+(165:1.2)$) {$+$};
\draw ($(3.3,0)+(20:\rR)$) -- ++(20:1.0);
\node at ($(3.3,0)+(20:2.4)$) {$3^+$};

\end{tikzpicture}\\
& 
\begin{tikzpicture}[baseline=-0.5ex,scale=0.62,thick,line cap=round,line join=round]
\def\rL{0.8}
\def\rR{0.8}

\draw[fill=gray!20] (0,0) circle (\rL);
\draw[fill=gray!20] (3.3,0) circle (\rR);
\node at (0,0) {$A_L$};
\node at (3.3,0) {$A_R$};

\draw (\rL,0) -- (3.3-\rR,0);

\draw (140:\rL) -- (140:1.48);
\node at (135:1.72) {$\hat{1}^-$};


\draw (160:\rL) -- (160:1.58);
\node at (160:2) {$7^+$};

\draw (200:\rL) -- (200:1.58);
\node at (200:2) {$6^+$};

\draw (225:\rL) -- (225:1.58);
\node at (226:2) {$5^-$};

\draw ($(3.3,0)+(40:\rR)$) -- ++(40:0.88);
\node at ($(3.3,0)+(40:2.2)$) {$\hat{2}^-$};

\draw ($(3.3,0)+(0:\rR)$) -- ++(0:0.88);
\node at ($(3.3,0)+(0:2.3)$) {$3^+$};

\draw ($(3.3,0)+(-40:\rR)$) -- ++(-40:0.88);
\node at ($(3.3,0)+(-40:2.3)$) {$4^+$};

\node at ($(0,0)+(15:1.2)$) {$+$};
\node at ($(3.3,0)+(165:1.2)$) {$-$};

\end{tikzpicture}
\qquad
\begin{tikzpicture}[baseline=-0.5ex,scale=0.62,thick,line cap=round,line join=round]
\def\rL{0.8}
\def\rR{0.8}

\draw[fill=gray!20] (0,0) circle (\rL);
\draw[fill=gray!20] (3.3,0) circle (\rR);
\node at (0,0) {$A_L$};
\node at (3.3,0) {$A_R$};

\draw (\rL,0) -- (3.3-\rR,0);

\draw (140:\rL) -- (140:1.48);
\node at (135:1.72) {$\hat{1}^-$};


\draw (160:\rL) -- (160:1.58);
\node at (160:2) {$7^+$};

\draw (180:\rL) -- (180:1.58);
\node at (180:2) {$6^+$};

\draw (200:\rL) -- (200:1.58);
\node at (200:2) {$5^-$};

\draw (225:\rL) -- (225:1.58);
\node at (226:2) {$4^+$};

\draw ($(3.3,0)+(40:\rR)$) -- ++(40:0.88);
\node at ($(3.3,0)+(40:2.2)$) {$\hat{2}^-$};

\draw ($(3.3,0)+(-40:\rR)$) -- ++(-40:0.88);
\node at ($(3.3,0)+(-40:2.3)$) {$3^+$};

\node at ($(0,0)+(15:1.2)$) {$-$};
\node at ($(3.3,0)+(165:1.2)$) {$+$};

\end{tikzpicture}
\end{aligned}
\end{equation*}

For the first three terms, the strict limits are the corresponding lower-point diagrams multiplied by
\begin{equation}
    \frac{2\, s_{1,2,P}\langle 5|p_1+P |2]}{\sqb{1}{2} \ab{P}{1} \ab{5}{P} (s_{1,2}+s_{1,2,P})}, \qquad \frac{2 \sqb{P}{2}}{\ab{1}{P}\sqb{1}{2}}, \qquad  \frac{2\ab{1}{5}}{\ab{1}{P}\ab{5}{P}},
\end{equation}
Although the first term is organized differently from Eq.~\eqref{eq: NMHV example term1}, factoring out the common lower-point diagram gives the same coefficient.  The first two terms therefore combine into the prefactor \eqref{eq: combine factor}.

\begin{samepage}
The fourth term contains a six-point NMHV subamplitude.  Using Eq.~\eqref{eq: 6-pt NMHV general}, its prefactor is
\begin{equation}
    \frac{2 \ab{1}{5} }{\ab{1}{P}\ab{5}{P}}-\frac{2\ab{1}{\hat I}\sqb{\hat I}{4}\ab{4}{5}}{\ab{1}{P}\ab{5}{P}(s_{1,\hat I}+s_{4,5})}= \frac{2 \ab{1}{5} }{\ab{1}{P}\ab{5}{P}}+\frac{2\langle1|p_5+P|4 ]\ab{4}{5}}{\ab{1}{P}\ab{5}{P}(s_{4,5,P}+s_{4,5})}
\end{equation}
\end{samepage}
The complete seven-point result is therefore
\begin{equation}
\begin{aligned}
    &A(1^-2^-3^+4^+5^-6^+7^+)\Big|^{(0)}_{\rm coll.} \xrightarrow[]{x=\frac12} \frac{2\ab{1}{5}}{\ab{1}{P}\ab{5}{P}} A(1^-2^-3^+4^+5^-P^+) \\
    &+ \frac{2\ab{1}{2}\langle 5|p_1+P|2]}{\ab{1}{P}\ab{5}{P}(s_{1,2}+s_{1,2,P})}\ \times \begin{tikzpicture}[baseline=-0.5ex,scale=0.62,thick,line cap=round,line join=round]
\def\rL{0.8}
\def\rR{0.8}

\draw[fill=gray!20] (0,0) circle (\rL);
\draw[fill=gray!20] (3.3,0) circle (\rR);
\node at (0,0) {$A_L$};
\node at (3.3,0) {$A_R$};

\draw (\rL,0) -- (3.3-\rR,0);

\draw (140:\rL) -- (140:1.48);
\node at (135:1.72) {$\hat{1}^-$};


\draw (225:\rL) -- (225:1.58);
\node at (226:2) {$P^+$};

\draw ($(3.3,0)+(40:\rR)$) -- ++(40:0.88);
\node at ($(3.3,0)+(40:2.2)$) {$\hat{2}^-$};

\draw ($(3.3,0)+(-40:\rR)$) -- ++(-40:0.88);
\node at ($(3.3,0)+(-40:2.3)$) {$5^-$};

\draw ($(3.3,0)+(-20:\rR)$) -- ++(-20:0.88);
\node at ($(3.3,0)+(-20:2.3)$) {$4^+$};

\node at ($(0,0)+(15:1.2)$) {$-$};
\node at ($(3.3,0)+(165:1.2)$) {$+$};
\draw ($(3.3,0)+(20:\rR)$) -- ++(20:1.0);
\node at ($(3.3,0)+(20:2.4)$) {$3^+$};

\end{tikzpicture}\\
&+  \frac{2\ab{4}{5}\langle1|p_5+P|4 ]}{\ab{1}{P}\ab{5}{P}(s_{4,5}+s_{4,5,P})} \ \times \
\begin{tikzpicture}[baseline=-0.5ex,scale=0.62,thick,line cap=round,line join=round]
\def\rL{0.8}
\def\rR{0.8}

\draw[fill=gray!20] (0,0) circle (\rL);
\draw[fill=gray!20] (3.3,0) circle (\rR);
\node at (0,0) {$A_L$};
\node at (3.3,0) {$A_R$};

\draw (\rL,0) -- (3.3-\rR,0);

\draw (140:\rL) -- (140:1.48);
\node at (135:1.72) {$\hat{1}^-$};


\draw (160:\rL) -- (160:1.58);
\node at (160:2) {$P^+$};

\draw (200:\rL) -- (200:1.58);
\node at (200:2) {$5^-$};

\draw (225:\rL) -- (225:1.58);
\node at (226:2) {$4^+$};

\draw ($(3.3,0)+(40:\rR)$) -- ++(40:0.88);
\node at ($(3.3,0)+(40:2.2)$) {$\hat{2}^-$};

\draw ($(3.3,0)+(-40:\rR)$) -- ++(-40:0.88);
\node at ($(3.3,0)+(-40:2.3)$) {$3^+$};

\node at ($(0,0)+(15:1.2)$) {$-$};
\node at ($(3.3,0)+(165:1.2)$) {$+$};

\end{tikzpicture}
\end{aligned}
\end{equation}
For an $n$-point NMHV amplitude whose three negative-helicity legs are $1,2$, and $n\!-\!2$, the following diagrams require separate treatment:
\begin{equation*}
\begin{aligned}
&\begin{tikzpicture}[baseline=-0.5ex,scale=0.62,thick,line cap=round,line join=round]
\def\rL{0.8}
\def\rR{0.8}

\draw[fill=gray!20] (0,0) circle (\rL);
\draw[fill=gray!20] (3.3,0) circle (\rR);
\node at (0,0) {$A_L$};
\node at (3.3,0) {$A_R$};

\draw (\rL,0) -- (3.3-\rR,0);

\draw (140:\rL) -- (140:1.48);
\node at (135:1.72) {$\hat{1}^-$};


\draw (225:\rL) -- (225:1.58);
\node at (226:2) {$n^+$};

\draw ($(3.3,0)+(70:\rR)$) -- ++(70:0.80);
\node at ($(3.3,0)+(70:2.1)$) {$\hat{2}^-$};

\draw ($(3.3,0)+(50:\rR)$) -- ++(50:0.80);
\node at ($(3.3,0)+(50:2.1)$) {$3^+$};

\draw ($(3.3,0)+(-30:\rR)$) -- ++(-30:0.80);
\node at ($(3.3,0)+(-20:2.6)$) {$(n\!-\!3)^+$};

\draw ($(3.3,0)+(-50:\rR)$) -- ++(-50:0.80);
\node at ($(3.3,0)+(-35:2.6)$) {$(n\!-\!2)^-$};

\draw ($(3.3,0)+(-70:\rR)$) -- ++(-70:0.80);
\node at ($(3.3,0)+(-75:2.2)$) {$(n\!-\!1)^+$};

\node at ($(3.3,0)+(30:1)$) {$\cdot$};
\node at ($(3.3,0)+(7.5:1)$) {$\cdot$};
\node at ($(3.3,0)+(-15:1)$) {$\cdot$};

\node at ($(0,0)+(15:1.2)$) {$-$};
\node at ($(3.3,0)+(165:1.2)$) {$+$};

\end{tikzpicture}
\qquad
\begin{tikzpicture}[baseline=-0.5ex,scale=0.62,thick,line cap=round,line join=round]
\def\rL{0.8}
\def\rR{0.8}

\draw[fill=gray!20] (0,0) circle (\rL);
\draw[fill=gray!20] (3.3,0) circle (\rR);
\node at (0,0) {$A_L$};
\node at (3.3,0) {$A_R$};

\draw (\rL,0) -- (3.3-\rR,0);

\draw (140:\rL) -- (140:1.48);
\node at (135:1.72) {$\hat{1}^-$};


\draw (180:\rL) -- (180:1.58);
\node at (180:2) {$n^+$};

\draw (225:\rL) -- (225:1.58);
\node at (225:2.2) {$(n\!-\!1)^+$};

\draw ($(3.3,0)+(70:\rR)$) -- ++(70:0.80);
\node at ($(3.3,0)+(70:2.1)$) {$\hat{2}^-$};

\draw ($(3.3,0)+(50:\rR)$) -- ++(50:0.80);
\node at ($(3.3,0)+(50:2.1)$) {$3^+$};

\draw ($(3.3,0)+(-45:\rR)$) -- ++(-45:0.80);
\node at ($(3.3,0)+(-40:2.3)$) {$(n\!-\!3)^+$};

\draw ($(3.3,0)+(-70:\rR)$) -- ++(-70:0.80);
\node at ($(3.3,0)+(-75:2.2)$) {$(n\!-\!2)^-$};

\node at ($(3.3,0)+(30:1)$) {$\cdot$};
\node at ($(3.3,0)+(7.5:1)$) {$\cdot$};
\node at ($(3.3,0)+(-15:1)$) {$\cdot$};

\node at ($(0,0)+(15:1.2)$) {$-$};
\node at ($(3.3,0)+(165:1.2)$) {$+$};

\end{tikzpicture}\\
& 
\hspace{3cm}
\begin{tikzpicture}[baseline=-0.5ex,scale=0.62,thick,line cap=round,line join=round]
\def\rL{0.8}
\def\rR{0.8}

\draw[fill=gray!20] (0,0) circle (\rL);
\draw[fill=gray!20] (3.3,0) circle (\rR);
\node at (0,0) {$A_L$};
\node at (3.3,0) {$A_R$};

\draw (\rL,0) -- (3.3-\rR,0);

\draw (120:\rL) -- (120:1.48);
\node at (110:1.8) {$\hat{1}^-$};


\draw (140:\rL) -- (140:1.58);
\node at (140:2) {$n^+$};

\draw (160:\rL) -- (160:1.58);
\node at (175:2.5) {$(n\!-\!1)^+$};

\draw (190:\rL) -- (190:1.58);
\node at (195:2.5) {$(n\!-\!2)^-$};

 \node at (224:1.00) {$\cdot$};
 \node at (208:1.00) {$\cdot$};

\draw (-120:\rL) -- (-120:1.58);
\node at (-120:2) {$4^+$};

\draw ($(3.3,0)+(40:\rR)$) -- ++(40:0.88);
\node at ($(3.3,0)+(40:2.2)$) {$\hat{2}^-$};

\draw ($(3.3,0)+(-40:\rR)$) -- ++(-40:0.88);
\node at ($(3.3,0)+(-40:2.3)$) {$3^+$};

\node at ($(0,0)+(15:1.2)$) {$-$};
\node at ($(3.3,0)+(165:1.2)$) {$+$};

\end{tikzpicture}
\end{aligned}
\end{equation*}
\begin{samepage}
The diagrams on the first line again combine at fixed lower-point amplitude with the prefactor \eqref{eq: combine factor}.  The final term contains an NMHV subamplitude of the same form, leading recursively to
\begin{equation}
\begin{aligned}
    &A(1^-2^-3^+\ldots(n\!-\!3)^+(n\!-\!2)^-(n\!-\!1)^+ n^+)\Big|^{(0)}_{\rm coll.} \xrightarrow[]{x=\frac12} \frac{2\ab{1}{n\!-\!2}}{\ab{1}{P}\ab{n-2}{P}} A(1^-2^-3^+\ldots(n\!-\!3)^+(n\!-\!2)^-P^+) \\ 
    &+ \frac{2\ab{1}{2}\langle n\!-\!2|p_1+P|2]}{\ab{1}{P}\ab{n\!-\!2}{P}(s_{1,2}+s_{1,2,P})}\ \times \begin{tikzpicture}[baseline=-0.5ex,scale=0.62,thick,line cap=round,line join=round]
\def\rL{0.8}
\def\rR{0.8}

\draw[fill=gray!20] (0,0) circle (\rL);
\draw[fill=gray!20] (3.3,0) circle (\rR);
\node at (0,0) {$A_L$};
\node at (3.3,0) {$A_R$};

\draw (\rL,0) -- (3.3-\rR,0);

\draw (140:\rL) -- (140:1.48);
\node at (135:1.72) {$\hat{1}^-$};


\draw (225:\rL) -- (225:1.58);
\node at (226:2) {$P^+$};

\draw ($(3.3,0)+(70:\rR)$) -- ++(70:0.80);
\node at ($(3.3,0)+(70:2.1)$) {$\hat{2}^-$};

\draw ($(3.3,0)+(50:\rR)$) -- ++(50:0.80);
\node at ($(3.3,0)+(50:2.1)$) {$3^+$};

\draw ($(3.3,0)+(-45:\rR)$) -- ++(-45:0.80);
\node at ($(3.3,0)+(-40:2.3)$) {$(n\!-\!3)^+$};

\draw ($(3.3,0)+(-70:\rR)$) -- ++(-70:0.80);
\node at ($(3.3,0)+(-75:2.2)$) {$(n\!-\!2)^-$};

\node at ($(3.3,0)+(30:1)$) {$\cdot$};
\node at ($(3.3,0)+(7.5:1)$) {$\cdot$};
\node at ($(3.3,0)+(-15:1)$) {$\cdot$};

\node at ($(0,0)+(15:1.2)$) {$-$};
\node at ($(3.3,0)+(165:1.2)$) {$+$};

\end{tikzpicture} \\
    &+\sum_{i=4}^{n-4} \frac{2\ab{1}{\hat{I}_i}\langle n\!-\!2|p_1+P|\hat{I}_i]}{\ab{1}{P}\ab{n\!-\!2}{P}(s_{1,\hat{I}_i}+s_{1,\hat{I}_i,P})}\ \times 
\begin{tikzpicture}[baseline=-0.5ex,scale=0.62,thick,line cap=round,line join=round]
\usetikzlibrary{calc}
\def\r{0.8}

\def\xA{0}
\def\xB{3.0}
\def\xC{6.0}
\def\xD{9.2}
\def\xE{12.0}

\draw[fill=gray!20] (\xA,0) circle (\r);
\draw[fill=gray!20] (\xB,0) circle (\r);
\draw[fill=gray!20] (\xC,0) circle (\r);
\draw[fill=gray!20] (\xD,0) circle (\r);
\draw[fill=gray!20] (\xE,0) circle (\r);

\draw (\xA+\r,0) -- (\xB-\r,0);
\draw (\xB+\r,0) -- (\xC-\r,0);
\draw (\xD+\r,0) -- (\xE-\r,0);

\draw (\xC+\r,0) -- (7,0);
\draw (8.2,0) -- (\xD-\r,0);

\node at ({(\xA+\xB)/2-0.4},0.33) {$-$};
\node at ({(\xA+\xB)/2+0.4},0.33) {$+$};

\node at ({(\xB+\xC)/2-0.45},0.33) {$-$};
\node at ({(\xB+\xC)/2+0.45},0.33) {$+$};

\node at ({(\xC+\xD)/2-0.5},0.33) {$-$};
\node at ({(\xC+\xD)/2+0.5},0.33) {$+$};

\node at ({(\xD+\xE)/2-0.4},0.33) {$-$};
\node at ({(\xD+\xE)/2+0.4},0.33) {$+$};

\node at ({(\xB+\xC)/2},-0.4) {$\hat{I}_i$};

\node at (7.62,0.02) {$\cdots$};

\draw ($(\xA,0)+(140:\r)$) -- ($(\xA,0)+(140:1.45)$);
\node at ($(\xA,0)+(140:1.8)$) {$\hat{1}^-$};

\draw ($(\xA,0)+(220:\r)$) -- ($(\xA,0)+(220:1.45)$);
\node at ($(\xA,0)+(220:1.9)$) {$P^+$};

\draw ($(\xB,0)+(240:\r)$) -- ++(240:0.95);
\node at ($(\xB,0)+(240:2.2)$) {$(n\!-\!2)^-$};

\draw ($(\xB,0)+(300:\r)$) -- ++(300:0.95);
\node at ($(\xB,0)+(300:2.1)$) {$i^+$};

\node at ($(\xB,0)+(250:1.0)$) {$\cdot$};
\node at ($(\xB,0)+(270:1.0)$) {$\cdot$};
\node at ($(\xB,0)+(290:1.0)$) {$\cdot$};

\draw ($(\xC,0)+(270:\r)$) -- ++(270:0.95);
\node at ($(\xC,0)+(270:2.0)$) {$(i\!-\!1)^+$};

\draw ($(\xD,0)+(270:\r)$) -- ++(270:0.95);
\node at ($(\xD,0)+(270:2.0)$) {$4^+$};

\draw ($(\xE,0)+(40:\r)$) -- ++(40:0.95);
\node at ($(\xE,0)+(40:2.4)$) {$\hat{2}^-$};

\draw ($(\xE,0)+(-35:\r)$) -- ++(-35:0.95);
\node at ($(\xE,0)+(-35:2.4)$) {$3^+$};

\end{tikzpicture} \\
&+  \frac{2\ab{n\!-\!3}{n\!-\!2}\langle1|p_{n-2}+P|n\!-\!3 ]}{\ab{1}{P}\ab{n\!-\!2}{P}(s_{n-3,n-2}+s_{n-3,n-2,P})} \ \times \
\begin{tikzpicture}[baseline=-0.5ex,scale=0.62,thick,line cap=round,line join=round]
\usetikzlibrary{calc}
\def\rL{0.8}
\def\rR{0.8}
\def\r{0.8}
\def\xA{0}
\def\xB{2.7}
\def\xC{5.9}
\def\xD{8.6}

\draw[fill=gray!20] (\xA,0) circle (\rL);
\draw[fill=gray!20] (\xB,0) circle (\rR);
\draw[fill=gray!20] (\xC,0) circle (\rR);
\draw[fill=gray!20] (\xD,0) circle (\rR);

\draw (\xA+\rL,0) -- (\xB-\rR,0);
\draw (\xC+\rR,0) -- (\xD-\rR,0);


\node at ({(\xA+\xB)/2-0.33},0.33) {$-$};
\node at ({(\xA+\xB)/2+0.33},0.33) {$+$};

\node at ({(\xB+\xC)/2-0.45},0.33) {$-$};
\node at ({(\xB+\xC)/2+0.45},0.33) {$+$};

\draw (\xB+\r,0) -- (3.7,0);
\draw (4.9,0) -- (\xC-\r,0);

\node at ({(\xC+\xD)/2-0.33},0.33) {$-$};
\node at ({(\xC+\xD)/2+0.33},0.33) {$+$};

\node at ({(\xB+\xC)/2},0) {$\cdots$};

\draw ($(\xA,0)+(140:\rL)$) -- ($(\xA,0)+(140:1.48)$);
\node at ($(\xA,0)+(135:1.72)$) {$\hat{1}^-$};

\draw ($(\xA,0)+(160:\rL)$) -- ($(\xA,0)+(160:1.58)$);
\node at ($(\xA,0)+(160:2)$) {$P^+$};

\draw ($(\xA,0)+(200:\rL)$) -- ($(\xA,0)+(200:1.58)$);
\node at ($(\xA,0)+(200:2.5)$) {$(n\!-\!2)^-$};

\draw ($(\xA,0)+(225:\rL)$) -- ($(\xA,0)+(225:1.58)$);
\node at ($(\xA,0)+(226:2.4)$) {$(n\!-\!3)^+$};

\node at ($(\xA,0)+(175:1.0)$) {$\cdot$};
\node at ($(\xA,0)+(188:1.00)$) {$\cdot$};

\draw ($(\xB,0)+(270:\rR)$) -- ++(270:0.95);
\node at ($(\xB,0)+(270:2.0)$) {$(n\!-\!4)^+$};

\draw ($(\xC,0)+(270:\rR)$) -- ++(270:0.95);
\node at ($(\xC,0)+(270:2.0)$) {$4^+$};

\draw ($(\xD,0)+(40:\rR)$) -- ++(40:0.88);
\node at ($(\xD,0)+(40:2.2)$) {$\hat{2}^-$};

\draw ($(\xD,0)+(-40:\rR)$) -- ++(-40:0.88);
\node at ($(\xD,0)+(-40:2.3)$) {$3^+$};

\end{tikzpicture}
\end{aligned}
\end{equation}
\end{samepage}

\section{Collinear legs with distinct polarizations}
\label{sec:distinct-polarizations}

So far we have taken the two collinear gluons to approach the same parent polarization. The construction extends directly to two distinct physical polarizations. This extension is useful both in general dimensions, where the parent gluon may carry any of the $D-2$ transverse polarizations, and in four dimensions, where it describes the ordered mixed-helicity limits $(-,+)$ and $(+,-)$.

\subsection{General-dimensional analysis}

Let $\varepsilon_P$ and $\varepsilon'_P$ be two physical polarization vectors of the null parent momentum,
\begin{equation}
 P\cdot\varepsilon_P=P\cdot\varepsilon'_P=0,
 \qquad
 \varepsilon_P^2=(\varepsilon'_P)^2=0.
 \label{eq:mixed-parent-polarizations}
\end{equation}
The two states are individually null, but their mixed contraction $\varepsilon_P\cdot\varepsilon'_P$ need not vanish. We assign $\varepsilon_P$ to leg $n-1$ and $\varepsilon'_P$ to leg $n$ and use the same null-rotation transport as in Eq.~\eqref{eq:pol_projection}:
\begin{equation}
\begin{aligned}
 \varepsilon_{n-1}^{\mu}(\varepsilon_P)
 &={\cal R}\!\left(-\frac{\epsilon}{x}k_\perp\right)
 \varepsilon_P^\mu,
 \\
 \varepsilon_n^{\mu}(\varepsilon'_P)
 &={\cal R}\!\left(\frac{\epsilon}{1-x}k_\perp\right)
 \varepsilon_P^{\prime\mu}.
\end{aligned}
\label{eq:mixed-pol-projection}
\end{equation}
Their expansions are
\begin{equation}
\begin{aligned}
 \varepsilon_{n-1}^{\mu}
 &=\varepsilon_P^\mu
 -\epsilon\frac{\varepsilon_P\cdot K}{xP\cdot K}k_\perp^\mu
 +\epsilon\frac{\varepsilon_P\cdot k_\perp}{xP\cdot K}K^\mu
 -\epsilon^2\frac{k_\perp^2(\varepsilon_P\cdot K)}
 {2x^2(P\cdot K)^2}K^\mu,
 \\
 \varepsilon_n^{\mu}
 &=\varepsilon_P^{\prime\mu}
 +\epsilon\frac{\varepsilon'_P\cdot K}{(1-x)P\cdot K}k_\perp^\mu
 -\epsilon\frac{\varepsilon'_P\cdot k_\perp}{(1-x)P\cdot K}K^\mu
 -\epsilon^2\frac{k_\perp^2(\varepsilon'_P\cdot K)}
 {2(1-x)^2(P\cdot K)^2}K^\mu.
\end{aligned}
\label{eq:mixed-pol-expansion}
\end{equation}

For the ordered mixed-polarization limit we continue the three-point numerator along Eq.~\eqref{eq:mixed-pol-projection}. Substitution into the factorization channel gives the leading term immediately:
\begin{equation}
\begin{aligned}
 A_n(\varepsilon_P,\varepsilon'_P)
 \Big|^{(-1)}_{\rm coll.}
 ={}&-\frac{2}{\epsilon k_\perp^2}\Big[
 x(\varepsilon'_P\cdot k_\perp)\varepsilon_P^\mu
 +(1-x)(\varepsilon_P\cdot k_\perp)\varepsilon_P^{\prime\mu}
 \\
 &\hspace{31mm}
 -x(1-x)(\varepsilon_P\cdot\varepsilon'_P)k_\perp^\mu
 \Big]A_\mu(1,2,\ldots,n-2,P).
\end{aligned}
\label{eq:mixed-leading-D}
\end{equation}
The last term originates from the third tensor structure of the three-point vertex. It vanishes when the two chosen transverse polarization states are orthogonal.  In four dimensions it is instead essential for opposite helicities; Appendix~\ref{app:four-dimensional-reduction} shows how it combines with the first two terms to produce the standard mixed-helicity splitting amplitudes.

At subleading order, define the pole-free hard contribution by
\begin{equation}
 \mathcal H_n(\varepsilon_P,\varepsilon'_P)
 :=\left(A_n-\frac{1}{s_{n-1,n}}
 \mathop{\rm Res}_{s_{n-1,n}=0}A_n\right)
 \bigg|_{\substack{p_{n-1}=xP,\;p_n=(1-x)P\\
 \varepsilon_{n-1}=\varepsilon_P,\;
 \varepsilon_n=\varepsilon'_P}}.
\label{eq:mixed-hard-D}
\end{equation}
For distinct polarizations, the subtraction in Eq.~\eqref{eq:mixed-hard-D} is implemented by the modified recursion of Section~\ref{subsec:modified-bg}. Namely, $\mathcal H_n(\varepsilon_P,\varepsilon'_P)$ is the strict value of the channel-deleted amplitude, whereas every finite term generated by continuing the $(n{-}1,n)$ factorization block away from the factorization surface is retained in the explicit pole contribution below. Appendix~\ref{app:modified-bg} proves that this separation loses no strict hard contribution. A direct expansion of the pole term then yields
\begin{equation}
\begin{aligned}
 A_n(\varepsilon_P,\varepsilon'_P)
 \Big|^{(0)}_{\rm coll.}
={}&\mathcal H_n(\varepsilon_P,\varepsilon'_P)
 \\
 &+\Bigg[
 \frac{\varepsilon_P\cdot K}{xP\cdot K}\varepsilon_P^{\prime\mu}
 -\frac{\varepsilon'_P\cdot K}{(1-x)P\cdot K}\varepsilon_P^\mu
 -\frac{2x-1}{P\cdot K}
 (\varepsilon_P\cdot\varepsilon'_P)K^\mu
 \Bigg]
 A_\mu(1,2,\ldots,n-2,P).
\end{aligned}
\label{eq:mixed-subleading-D}
\end{equation}
The first two terms in brackets are required by the null-rotation transport of the two independent parent states; the last receives contribution from the explicit $O(\epsilon^2)K^\mu$ numerator. 

Under the four-dimensional specialization in Eqs.~\eqref{eq: 4d kperp} and \eqref{eq: 4d K}, the parent polarizations may carry arbitrary and independent reference spinors.  Every occurrence of $K$ in Eq.~\eqref{eq:mixed-subleading-D} is invariant under $K\to cK$, so the choice $K=r$ introduces no additional normalization factors.  With the convention fixed above, Eq.~\eqref{eq:mixed-subleading-D} reproduces the direct four-dimensional mixed-helicity expansion independently of those gauge references.

For the symmetric momentum split, Eq.~\eqref{eq:mixed-subleading-D} simplifies to
\begin{equation}
\begin{aligned}
 A_n(\varepsilon_P,\varepsilon'_P)
 \Big|^{(0)}_{\rm coll.}
 \xrightarrow{x=1/2}
 \mathcal H_n(\varepsilon_P,\varepsilon'_P)
 +\frac{2}{P\cdot K}
 \left[(\varepsilon_P\cdot K)\varepsilon_P^{\prime\mu}
 -(\varepsilon'_P\cdot K)\varepsilon_P^\mu\right]
 A_\mu(1,2,\ldots,n-2,P).
\end{aligned}
\label{eq:mixed-half-D}
\end{equation}
The explicit pole correction is antisymmetric under $\varepsilon_P\leftrightarrow\varepsilon'_P$.  Consequently, for the polarization-symmetrized combination
\begin{equation}
 A_n^{\rm sym}(\varepsilon_P,\varepsilon'_P)
 :=A_n(\varepsilon_P,\varepsilon'_P)
 +A_n(\varepsilon'_P,\varepsilon_P),
\label{eq:mixed-sym-D}
\end{equation}
where the momenta and color ordering are held fixed while the two parent polarizations assigned to legs $n-1$ and $n$ are exchanged, one finds
\begin{equation}
 A_n^{\rm sym}(\varepsilon_P,\varepsilon'_P)
 \Big|^{(0)}_{\rm coll.}
 \xrightarrow{x=1/2}
 \mathcal H_n(\varepsilon_P,\varepsilon'_P)
 +\mathcal H_n(\varepsilon'_P,\varepsilon_P).
\label{eq:mixed-sym-half-D}
\end{equation}
Thus the symmetric strict coefficient is manifestly \textbf{independent} of both $K$ and $k_\perp$.  Upon reduction to four dimensions, this becomes the cancellation of the $r$ dependence between the $(-,+)$ and $(+,-)$ components, as shown below.

\subsection{Four-dimensional mixed-helicity limits}

We now consider the case in which the two collinear legs have distinct helicities in four dimensions. Using the shorthand notation $c=\sqrt{x}$ and $s=\sqrt{1-x}$, we consider an MHV amplitude in which an external leg $i$ and a collinear leg carry negative helicity. The external legs adjacent to the collinear pair are $n-2$ and $1$. A direct expansion of the Parke--Taylor formula gives
\begin{equation}
\begin{aligned}
 &A(\ldots i^-\ldots n-2,(n-1)^-,n^+)
 =\left[\frac{c^3}{s}\frac{1}{\epsilon\ab{P}{r}}
 +S^{(0)}_{-+}\right]
 A(\ldots i^-\ldots n-2,P^-)+O(\epsilon),
 \\
 &A(\ldots i^-\ldots n-2,(n-1)^+,n^-)
 =\left[\frac{s^3}{c}\frac{1}{\epsilon\ab{P}{r}}
 +S^{(0)}_{+-}\right]
 A(\ldots i^-\ldots n-2,P^-)+O(\epsilon),
\end{aligned}
\label{eq:mixed-MHV-expansion}
\end{equation}
where
\begin{equation}
\begin{aligned}
 S^{(0)}_{-+}
 &=\frac{c^3}{s\ab{P}{r}}
 \left[
 -4\frac{s}{c}\frac{\ab{i}{r}}{\ab{i}{P}}
 +\frac{s}{c}\frac{\ab{n-2}{r}}{\ab{n-2}{P}}
 -\frac{c}{s}\frac{\ab{r}{1}}{\ab{P}{1}}
 \right],
 \\
 S^{(0)}_{+-}
 &=\frac{s^3}{c\ab{P}{r}}
 \left[
 +4\frac{c}{s}\frac{\ab{i}{r}}{\ab{i}{P}}
 +\frac{s}{c}\frac{\ab{n-2}{r}}{\ab{n-2}{P}}
 -\frac{c}{s}\frac{\ab{r}{1}}{\ab{P}{1}}
 \right].
\end{aligned}
\label{eq:mixed-MHV-seeds}
\end{equation}
These formulas include the endpoint case $i=n-2$.  In that case one simply identifies the two labels and combines the two terms proportional to $\langle n-2\,r\rangle/\langle n-2\,P\rangle$; no separate terminal rule is required. Their parity conjugates follow by exchanging angle and square brackets and reversing all helicities.

At $x=1/2$, the two ordered finite coefficients become
\begin{equation}
\begin{aligned}
 S^{(0)}_{-+}\big|_{x=1/2}
 &=\frac{1}{2\ab{P}{r}}
 \left[-4\frac{\ab{i}{r}}{\ab{i}{P}}
 +\frac{\ab{n-2}{r}}{\ab{n-2}{P}}
 -\frac{\ab{r}{1}}{\ab{P}{1}}\right],
 \\
 S^{(0)}_{+-}\big|_{x=1/2}
 &=\frac{1}{2\ab{P}{r}}
 \left[+4\frac{\ab{i}{r}}{\ab{i}{P}}
 +\frac{\ab{n-2}{r}}{\ab{n-2}{P}}
 -\frac{\ab{r}{1}}{\ab{P}{1}}\right].
\end{aligned}
\label{eq:mixed-MHV-half}
\end{equation}
Each ordered helicity assignment depends on the reference spinor $r$. In the helicity-symmetrized sum, the terms involving the negative-helicity hard leg cancel, and Schouten's identity gives
\begin{equation}
 \left(S^{(0)}_{-+}+S^{(0)}_{+-}\right)\big|_{x=1/2}
 =\frac{\ab{n-2}{1}}{\ab{n-2}{P}\ab{P}{1}},
\label{eq:mixed-MHV-sym}
\end{equation}
which is independent of $r$. Thus the cancellation is the four-dimensional counterpart of the antisymmetry displayed in Eq.~\eqref{eq:mixed-half-D}.

Finally, the BCFW construction of Section~\ref{sec:four-dimensional-collinear} requires no structural modification. If the two collinear legs occur in different subamplitudes, one takes the same strict hard limit as before. If they occur in the same subamplitude, the collinear operation is applied recursively to that subamplitude. The only new terminal data are the mixed MHV and $\overline{\rm MHV}$ seeds in Eq.~\eqref{eq:mixed-MHV-seeds}. One may retain either ordered component, $(-,+)$ or $(+,-)$, in which case its $x=1/2$ coefficient generally depends on $r$, or sum the two assignments, in which case the reference-spinor dependence cancels.

\section{Einstein--Yang--Mills amplitudes and string corrections from collinear data}
\label{sec:eym-from-collinear}

Mixed open--closed disk amplitudes give a complementary interpretation of the strict collinear coefficients computed above.  A closed-string insertion in the interior of the disk can be represented, after contour deformation and monodromy reduction, by two additional open-string insertions on its boundary \cite{Stieberger:0907,Stieberger:2015kia}.  The resulting exact relation is
\begin{align}
A(1,2,\ldots,n{-}2;q_1,q_2)
&=(-1)^n e^{-\pi i(s_{1,n}+s_{2,n-1})}
 \sum_{l=2}^{n-2}(-1)^l\sin(\pi s_{l,n-1})
 e^{\pi i(-1)^l s_{l,n-1}}
 \nonumber\\[-1mm]
&\quad\times
 \sum_{\rho\in\{OP(\alpha,\beta^t),l\}}
 e^{\pi i\sum_{k=1}^{\lfloor (n-3)/2\rfloor}\tau_{2k+1}(\rho)}
 {\cal S}(\rho)\,A(1,\rho,n-1,n).
\label{eq:mixed-disk-exact}
\end{align}
Here, only in the string-theory formulas of this section, $s_{ij}:=2\alpha' k_i\!\cdot k_j$ denotes a dimensionless Mandelstam invariant.  For each $l$, $\alpha=\{2,\ldots,l-1\}$ and $\beta=\{l+1,\ldots,n-2\}$, while $OP(\alpha,\beta^t)$ denotes the order-preserving merger of $\alpha$ with the reversed word $\beta^t$.  The functions ${\cal S}(\rho)$ and $\tau_{2k+1}(\rho)$ collect the monodromy phases:
\be\label{kernel}
\Sc(\rho)\equiv \Sc[ \rho(2,\ldots,n-2) \, ] = \prod_{i=2}^{n-2}\prod_{j=i+1}^{n-2}
\exp\lf\{  \pi i\;\Theta(\rho^{-1}(i)-\rho^{-1}(j)) \ s_{i,j} \ri\}\ ,
\ee
with $\Theta$ denoted the Heaviside step function and
\be\label{TAU}
\tau_i(\rho)=\begin{cases}
{\rm sign}(\rho^{-1}(i)-\rho^{-1}(i+1))\ (s_{i,n-1}+s_{i+1,n-1})\ ,&3\leq i\leq n-3\ ,\\
s_{n-2,n-1}\ ,&i=n\!-\!2\ .
\end{cases}
\ee

The open super string amplitudes on the right-hand side of equation \eqref{eq:mixed-disk-exact} can in turn be decomposed into Yang--Mills amplitudes in a BCJ basis~\cite{Mafra:2011nv,Mafra:2011nw,Schlotterer:2012ny},
\begin{equation}
 \vec A_{\rm open}(\alpha')
 =\mathbf F(\alpha')\,\vec A_{\rm YM},
 \qquad
 \mathbf F
 =\mathbf P\,\mathbf Q:
 \exp\!\left(\sum_{k\geq1}\zeta_{2k+1}\mathbf M_{2k+1}\right):\, .
\label{eq:open-to-YM-F}
\end{equation}
The entries of the period matrix $\mathbf F$ are the ordered open-string $F$-integrals.  They, and hence the complete $\alpha'$ expansion of Eq.~\eqref{eq:mixed-disk-exact}, can be generated recursively from the Drinfeld associator \cite{Broedel:2013aza}.  Substituting Eq.~\eqref{eq:open-to-YM-F} into Eq.~\eqref{eq:mixed-disk-exact} and reducing to a BCJ basis gives
\begin{equation}
 A(1,2,\ldots,n{-}2;q_1,q_2)
 =\sum_{\rho\in S_{n-3}}
 c_\rho(\epsilon,\alpha')
 A_{{\rm YM},\rho}(\epsilon),
 \qquad
 A_{{\rm YM},\rho}:=A_{\rm YM}(1,\rho,n-1,n),
\label{eq:disk-relation-summary}
\end{equation}
where the coefficients $c_\rho$ now include both the monodromy kernel in Eq.~\eqref{eq:mixed-disk-exact} and the $F$-integrals in Eq.~\eqref{eq:open-to-YM-F}.

In the field-theory collinear limit, the closed-string state becomes a graviton and the two effective boundary states become a pair of collinear gluons.  This realizes a one-graviton Einstein--Yang--Mills amplitude as a weighted combination of subleading Yang--Mills collinear limits \cite{Stieberger:2014hba,Stieberger:2015kia_PLB,Stieberger:2016lag}.  More precisely, expanding
\begin{equation}
 c_\rho=c_{\rho,0}+\epsilon c_{\rho,1}+O(\epsilon^2),
 \qquad
 A_{{\rm YM},\rho}
 =A_{{\rm YM},\rho}\Big|^{(-1)}_{\rm coll.}
 +A_{{\rm YM},\rho}\Big|^{(0)}_{\rm coll.}
 +O(\epsilon),
\label{eq:disk-collinear-summary}
\end{equation}
the strict collinear projection freezes the string kernel on the collinear surface before selecting the finite Yang--Mills coefficient:
\begin{equation}
 \left.A(1,2,\ldots,n{-}2;q_1,q_2)\right|_{\rm strict,\,\epsilon^0}
 :=\sum_{\rho\in S_{n-3}}c_{\rho,0}(\alpha')
 A_{{\rm YM},\rho}\Big|^{(0)}_{\rm coll.}.
\label{eq:strict-disk-summary}
\end{equation}
Note that it does not include the off-collinear pole-mixing term $\sum_\rho \epsilon c_{\rho,1}A_{{\rm YM},\rho}\big|^{(-1)}_{\rm coll.}$~\cite{DongStieberger:toappear}.  At $x=1/2$ the equal-helicity terms $A_{{\rm YM},\rho}\big|^{(0)}_{\rm coll.}$ are individually independent of the reference spinor.

The companion paper \cite{DongStieberger:toappear} studies Eq.~\eqref{eq:strict-disk-summary} in the direction from gravity to gauge theory.  It shows that the $(n-3)!$ strict subleading Yang--Mills coefficients can be determined from EYM amplitudes, their higher-derivative open-string corrections, and the complementary homogeneous BCJ-like relations.  The counting is governed by unsigned Stirling numbers of the first kind: ordinary one-graviton EYM amplitudes occupy the sector $\genfrac{[}{]}{0pt}{}{n-3}{1}=(n-4)!$, higher-derivative EYM data occupy the odd sectors $\genfrac{[}{]}{0pt}{}{n-3}{2j+1}$, and the homogeneous relations occupy the even sectors $\genfrac{[}{]}{0pt}{}{n-3}{2j}$.  The even and odd sectors each sum to $\frac12(n-3)!$ and together form a square system of full rank at generic kinematics, as verified there through nine points.

In the present paper, the same relation becomes a direct computational tool. Since the recursions of Sections~\ref{sec:general-dimensional-collinear} and \ref{sec:four-dimensional-collinear} compute every $A_{{\rm YM},\rho}\big|^{(0)}_{\rm coll.}$ directly, Eq.~\eqref{eq:strict-disk-summary} constructs the EYM amplitude and each of its string corrections.  If
\begin{equation}
\left.A(1,2,\ldots,n{-}2;q_1,q_2)\right|_{\rm strict,\,\epsilon^0}
 =\sum_{m\geq0}(\alpha')^{m+1} A_{\rm EYM}^{(m)},
 \qquad
 A_{\rm EYM}^{(m)}=\sum_{\rho\in S_{n-3}}
 G_\rho^{(m)}A_{{\rm YM},\rho}\Big|^{(0)}_{\rm coll.},
\label{eq:disk-amplitude-expansion}
\end{equation}
then $A_{\rm EYM}^{(0)}$ gives the ordinary EYM amplitude, up to the standard overall normalization, and the higher $A_{\rm EYM}^{(m)}$ give its successive string corrections.  Thus no separate EYM recursion is required: the desired gravitational quantity is assembled from ordinary color-ordered Yang--Mills amplitudes and kernels that are calculable from the Drinfeld associator.

There is an important distinction between computability and independence in this statement.  
The permutations of the expansion in~\eqref{eq:disk-amplitude-expansion} gives rise to the following set of equations:
\begin{equation}
\left.A(1,\sigma;q_1,q_2)\right|_{\rm strict\,,\epsilon^0}
=\sum_{m\geq0}(\alpha')^{m+1} \sum_{\rho\in S_{n-3}}
G_{\rho,\sigma}^{(m)}A_{{\rm YM},\rho}\Big|^{(0)}_{\rm coll.},
\quad
\mathbf G^{(m)}:=G_{\rho,\sigma}^{(m)}\,.
\end{equation}
Here we have introduced the shorthand notation $\mathbf G^{(m)}$ for the corresponding coefficient matrix.
The analysis of Ref.~\cite{DongStieberger:toappear} shows that the real part of the odd-order coefficient matrices ${\rm Re}\, \mathbf G^{(2j+1)}$ generate new independent EYM-type sectors, whereas the imaginary part even-order coefficient matrices ${\rm Im}\,\mathbf G^{(2j)}$ yield homogeneous BCJ-like relations, while its real part does not introduce additional independent EYM-type amplitudes.  This does not mean that the latter corrections vanish or cannot be obtained.  Equations \eqref{eq:mixed-disk-exact}--\eqref{eq:disk-amplitude-expansion} compute their contributions just as directly.  Rather, the available examples suggest that, the real part of the even-order contributions lie in the module generated by the real part of odd-order sectors over the ring of Mandelstam polynomials: they can be written as Mandelstam-polynomial multiples of lower odd-order data.  They therefore contain calculable string corrections but no new independent equations.  

We illustrate this construction with a few explicit relations at the symmetric point $x=1/2$.  For $m=n-2$ hard gluons, define the shorthand
\begin{equation}
 A_{n}(1,\rho,n)_{\rm sub}
 :=
 \left.
 A_{\rm YM}(1,\rho,n-1,n)
 \right|_{\rm coll.}^{(0)},
 \qquad
 p_{n-1}=p_n=\frac{P}{2}.
 \label{eq:Amp2-sub-shorthand}
\end{equation}
In the formulas below we return to the field-theory convention $s_I=(\sum_{i\in I}p_i)^2$; the appropriate power of $\alpha'$ has already been extracted from each coefficient.  We use the normalization inherited directly from the disk relation.  In this normalization the coefficient at overall order $\alpha'$ gives the ordinary EYM amplitude, while the coefficient at order $\alpha'^3$ gives its first $F^4$ correction.

At five points, the ordinary one-graviton EYM amplitude follows from
\begin{equation}
 A_{\rm EYM}(1,2,3;P)
 =\frac{\pi}{2}\left\{
 s_{12}A_5(1,2,3,4,5)_{\rm sub}
 +(s_{12}+s_{23})A_5(1,3,2,4,5)_{\rm sub}
 \right\}.
\label{eq:EYM-five-from-collinear}
\end{equation}
At six points, the corresponding relation is
\begin{align}
 A_{\rm EYM}(1,2,3,4;P)
 =\frac{\pi}{2}\Big\{&
 (s_{12}+s_{13}+s_{23})A_6(1,2,3,4,5,6)_{\rm sub}
 \nonumber\\
 &-(s_{12}+s_{23}+s_{24})A_6(1,4,3,2,5,6)_{\rm sub}
 \nonumber\\
 &+(s_{13}+s_{23}+s_{34})
 \big(A_6(1,2,4,3,5,6)_{\rm sub}+A_6(1,4,2,3,5,6)_{\rm sub}\big)
 \Big\}.
\label{eq:EYM-six-from-collinear}
\end{align}

The same six-point data also determine the first genuinely new higher-derivative contribution, which we denote by $A_{\rm EYM}^{F^4}$.  The $F^4$ correction cannot occur in the five-point relation \eqref{eq:EYM-five-from-collinear}, since an $F^4$ interaction requires at least four external gluons.  It first appears at $n=6$ in the unfused Yang--Mills labeling: after legs $5$ and $6$ are fused, the corresponding EYM amplitude contains precisely the minimal configuration of four gluons and one graviton.  Collecting the result by color ordering gives
\begin{align}
 A_{\rm EYM}^{F^4}(1,2,3,4;P)
 =\frac{\pi \zeta_2}{2}\Big\{&
 s_{12}s_{123}(s_{12}-s_{123})A_6(1,2,3,4,5,6)_{\rm sub}
 \nonumber\\
 &+s_{12}(s_{34}+s_{123})(s_{12}-s_{34}-s_{123})
 A_6(1,2,4,3,5,6)_{\rm sub}
 \nonumber\\
 &+s_{123}(s_{12}+s_{23})(s_{12}+s_{23}-s_{123})
 A_6(1,3,2,4,5,6)_{\rm sub}
 \nonumber\\
 &-(s_{12}-s_{34}+s_{234})(-s_{23}+s_{123}+s_{234})
 \nonumber\\[-1mm]
 &\hspace{13mm}\times(-s_{12}-s_{23}+s_{34}+s_{123})
 A_6(1,4,3,2,5,6)_{\rm sub}
 \nonumber\\
 &-(s_{12}+s_{23}-s_{123})(s_{23}+s_{34}-s_{123}-s_{234})
 \nonumber\\[-1mm]
 &\hspace{13mm}\times(s_{12}-s_{34}+s_{234})
 A_6(1,3,4,2,5,6)_{\rm sub}
 \nonumber\\
 &+(s_{12}-s_{34}-s_{123})(s_{12}-s_{23}-s_{34}+s_{234})
 \nonumber\\[-1mm]
 &\hspace{13mm}\times(-s_{23}+s_{123}+s_{234})
 A_6(1,4,2,3,5,6)_{\rm sub}
 \Big\}.
\label{eq:F4-six-from-collinear}
\end{align}
The RHS of~\eqref{eq:F4-six-from-collinear} contains the cubic Mandelstam polynomials obtained from the $\alpha'^3$ coefficient. Let us also spell out explicitly the seven-point relation for the leading EYM amplitude:
\begin{align}
 A_{\rm EYM}(1,2,3,4,5;P)
 =\frac{\pi}{2}\Big\{&
 (s_{12}+s_{23}+s_{24}+s_{25})A_7(1,5,4,3,2,6,7)_{\rm sub}
 \nonumber\\
 &+(s_{12}+s_{13}+s_{14}+s_{23}+s_{24}+s_{34})
 A_7(1,2,3,4,5,6,7)_{\rm sub}
 \nonumber\\
 &-(s_{13}+s_{23}+s_{34}+s_{35})
 \nonumber\\[-1mm]
 &\hspace{13mm}\times\big[A_7(1,2,5,4,3,6,7)_{\rm sub}
 +A_7(1,5,2,4,3,6,7)_{\rm sub}
 \nonumber\\[-1mm]
 &\hspace{43mm}
 +A_7(1,5,4,2,3,6,7)_{\rm sub}\big]
 \nonumber\\
 &+(s_{14}+s_{24}+s_{34}+s_{45})
 \nonumber\\[-1mm]
 &\hspace{13mm}\times\big[A_7(1,2,3,5,4,6,7)_{\rm sub}
 +A_7(1,2,5,3,4,6,7)_{\rm sub}
 \nonumber\\[-1mm]
 &\hspace{43mm}
 +A_7(1,5,2,3,4,6,7)_{\rm sub}\big]
 \Big\}.
\label{eq:EYM-seven-from-collinear}
\end{align}
These examples exhibit the disk relation as a direct map from collinear Yang--Mills data to mixed gauge--gravity quantities.  Once the strict coefficients $A_{n}(\cdots,n-1,n)_{\rm sub}$ are generated by recursion developed in this paper, the ordinary EYM amplitudes and their higher-derivative corrections follow as polynomial linear combinations.

\paragraph{Polarization content.}
We finally spell out the state selected by the two collinear gluons.  The polarizations of legs $n-1$ and $n$ provide the two vector factors of the closed-state polarization tensor.  In four dimensions, equal helicities therefore select the two graviton helicities,
\begin{equation}
 \big((n-1)^+,n^+\big)\ \longleftrightarrow\ h^{+2},
 \qquad
 \big((n-1)^-,n^-\big)\ \longleftrightarrow\ h^{-2}.
 \label{eq:closed-state-equal-helicity}
\end{equation}
These are the choices implicit in the one-graviton EYM relations displayed above.  The dilaton is selected instead by the symmetric sum of the two opposite-helicity assignments~\cite{Stieberger:2014hba}.  Up to state-normalization and metric-sign conventions,
\begin{equation}
 A_{n,\phi}(1,\rho,n-1,n)_{\rm sub}
 \ \propto\
 A_{n}(1,\rho,(n-1)^+,n^-)_{\rm sub}
 +A_{n}(1,\rho,(n-1)^-,n^+)_{\rm sub}.
 \label{eq:dilaton-from-opposite-helicity}
\end{equation}
This is precisely the polarization-symmetrized combination of Section~\ref{sec:distinct-polarizations}; at $x=1/2$ its dependence on the reference spinor $r$ cancels.  The same disk kernels then give the one-dilaton amplitude and its higher-derivative corrections in place of the one-graviton EYM amplitude.  

The corresponding $D$-dimensional statement is a trace over the physical transverse polarization space.  Let $\{\varepsilon_P^{(I)}\}$ be a real orthonormal basis of the $D-2$ physical states transverse to both $P$ and an auxiliary null vector $Q$, and define
\begin{equation}
 \Pi_{\mu\nu}(P,Q)
 :=\sum_{\text{physical states }I}
 \varepsilon_{P,\mu}^{(I)}\varepsilon_{P,\nu}^{(I)}
 =\eta_{\mu\nu}
 -\frac{P_\mu Q_\nu+Q_\mu P_\nu}{P\cdot Q},
 \qquad P^2=Q^2=0,
 \label{eq:transverse-polarization-projector}
\end{equation}
where an overall sign on the right-hand side follows the convention for the norm of a physical polarization.  To display the contraction explicitly, let $A_n^{\mu\nu}\big|_{\rm sub}$ denote the strict coefficient with the two parent polarization vectors stripped off, so that
\begin{equation}
 A_n(\ldots,\varepsilon_P,\varepsilon'_P)_{\rm sub}
 =\varepsilon_{P,\mu}\varepsilon'_{P,\nu}
 A_n^{\mu\nu}\Big|_{\rm sub}.
 \label{eq:polarization-stripped-subleading-amplitude}
\end{equation}
The normalized dilaton polarization and amplitude are then
\begin{equation}
 \begin{aligned}
 \varepsilon^{(\phi)}_{\mu\nu}
 &=\frac{1}{\sqrt{D-2}}\Pi_{\mu\nu}(P,Q),
 \\
 A_{n,\phi}\big|_{\rm sub}
 &=\frac{1}{\sqrt{D-2}}\Pi_{\mu\nu}(P,Q)
 A_n^{\mu\nu}\Big|_{\rm sub}
 \\
 &=\frac{1}{\sqrt{D-2}}\sum_{\text{physical states }I}
 A_{n}\big(\ldots,
 \varepsilon_P^{(I)},\varepsilon_P^{(I)}\big)_{\rm sub}.
 \end{aligned}
 \label{eq:D-dimensional-dilaton-trace}
\end{equation}
The $Q$-dependent terms in Eq.~\eqref{eq:transverse-polarization-projector} are proportional to the closed momentum in one index and drop from the complete gauge-invariant amplitude by the Ward identities.  The graviton is obtained from the symmetric traceless projection~\footnote{Equivalently, one may set $\varepsilon_P'=\varepsilon_P$, as in the previous sections.}
\begin{equation}
 \varepsilon^{(h)}_{\mu\nu}
 =\varepsilon_{P,(\mu}\varepsilon'_{P,\nu)}
 -\frac{\varepsilon_P\cdot\varepsilon'_P}{D-2}
 \Pi_{\mu\nu}(P,Q),
 \qquad
 \eta^{\mu\nu}\varepsilon^{(h)}_{\mu\nu}=0,
 \label{eq:D-dimensional-graviton-projection}
\end{equation}
whereas the transverse trace in Eq.~\eqref{eq:D-dimensional-dilaton-trace} selects the dilaton.  Thus a single pair of distinct transverse polarizations is not, in general, a pure dilaton state; the complete trace over the $D-2$ physical polarizations is required.

\section{Mathematica implementation}
\label{sec:mathematica-package}

The ancillary file \texttt{CollinearRecursion.wl} provides a self-contained Wolfram Language implementation of the covariant and four-dimensional constructions used in this paper. It represents momenta and polarizations by \texttt{p[i]} and \texttt{e[i]}, a second parent polarization by \texttt{ePrime[P]}, Lorentz products by \texttt{SP[a,b]}, and spinor brackets by \texttt{ab[i,j]} and \texttt{sb[i,j]}. The default polarization-reference spinor of \texttt{e[i]} is denoted by \texttt{eta[i]}, while the default reference of \texttt{ePrime[i]} is \texttt{etaPrime[i]}.  The package is loaded with
\begin{verbatim}
Get["CollinearRecursion.wl"];
\end{verbatim}
and has no external package dependencies.

The principal public functions are summarized in Table~\ref{tab:package-functions}.
\begin{table}[H]
\centering
\small
\renewcommand{\arraystretch}{1.2}
\begin{tabularx}{\textwidth}{@{}>{\raggedright\arraybackslash}p{0.34\textwidth}X@{}}
\hline
\textbf{Function} & \textbf{Purpose} \\
\hline
\texttt{YMAmplitudeD}\newline\texttt{[legs]} &
Color-ordered tree amplitude from covariant Berends--Giele recursion.\\
\texttt{CollinearSubleadingD}\newline\texttt{[legs, opts]} &
Strict $\epsilon^0$ adjacent-collinear coefficient from the modified
Berends--Giele recursion of Section~\ref{subsec:modified-bg}.\\
\texttt{YMAmplitude4D}\newline\texttt{[helicities, opts]} &
Four-dimensional tree amplitude from analytic BCFW recursion.\\
\texttt{CollinearSubleading4D}\newline\texttt{[helicities, opts]} &
Strict $\epsilon^0$ coefficient from the collinear-aware BCFW recursion,
without expanding the complete amplitude in $\epsilon$.\\
\texttt{MHVCollinearReplacement}\newline\texttt{[helicities, opts]} &
Closed MHV or $\overline{\rm MHV}$ terminal rule at generic $x$, including
equal- and mixed-helicity pairs.\\
\texttt{DTo4D}\newline\texttt{[expr, helicities, opts]} &
Reduction of a covariant expression to four-dimensional spinor-helicity
variables.\\
\texttt{ExpandScalarProducts}\newline\texttt{[expr]} &
Bilinear expansion of the symbolic Lorentz product \texttt{SP}.\\
\hline
\end{tabularx}
\caption{Main functions in the accompanying Mathematica package.}
\label{tab:package-functions}
\end{table}

For example, the covariant five-point coefficient with legs $4$ and $5$ collinear is obtained from
\begin{verbatim}
subD = CollinearSubleadingD[
  Range[5], CollinearLegs -> {4, 5},
  ParentLeg -> P, KVectorLabel -> K,
  MomentumFraction -> x];
\end{verbatim}
Setting \texttt{SymmetricCollinearLimit -> True} imposes $x=1/2$ before the recursion and avoids computing terms that vanish at the symmetric point. The polarization choices are controlled by
\begin{quote}
\ttfamily DistinctCollinearPolarizations -> True,\newline SymmetrizeCollinearPolarizations -> True.
\end{quote}
They respectively select the ordered pair $(\varepsilon_P,\varepsilon'_P)$ and the sum over its two assignments.

In four dimensions, helicities are specified by $-1$ and $+1$. A direct generic-$x$ computation is, for instance,
\begin{verbatim}
sub4 = CollinearSubleading4D[
  {-1, -1, 1, 1, 1},
  ParentLeg -> P, ReferenceLeg -> r,
  MomentumFraction -> x, ShiftLegs -> {1, 2}];
\end{verbatim}
The last two entries are the adjacent collinear legs. The option \texttt{ShiftLegs} fixes only the first BCFW bridge; lower subamplitudes choose valid shifts automatically. The production routine implements the two recursive cases described in Section~\ref{sec:four-dimensional-collinear} directly and does not invoke \texttt{Series} or first construct an $\epsilon$-dependent full amplitude.

The covariant output can be reduced to a chosen four-dimensional helicity component. Continuing the preceding example,
\begin{verbatim}
sub4FromD = DTo4D[
  subD, <|1 -> -1, 2 -> -1, 3 -> 1, P -> 1|>,
  LegLabels -> {1, 2, 3, P}, ParentLeg -> P,
  ReferenceLeg -> r, KVectorLabel -> K,
  PolarizationReferences -> <|P -> eta[P]|>,
  MomentumFraction -> x];
\end{verbatim}
The displayed choice \texttt{eta[P]} is also the default reference spinor for \texttt{e[P]}, so the option \texttt{PolarizationReferences} may be omitted in this example.
For distinct parent polarizations, the corresponding covariant expression is generated by
\begin{verbatim}
subDDistinct = CollinearSubleadingD[
  Range[5], CollinearLegs -> {4, 5},
  ParentLeg -> P, KVectorLabel -> K,
  TransverseVectorLabel -> kPerp,
  MomentumFraction -> x,
  DistinctCollinearPolarizations -> True];
\end{verbatim}
\begin{samepage}
Thus the option explicitly assigns the ordered pair $(\varepsilon_P,\varepsilon'_P)$ to legs $4$ and $5$.  For example, the component in which \texttt{e[P]} has negative helicity and \texttt{ePrime[P]} has positive helicity is obtained from
\begin{verbatim}
sub4MixedFromD = DTo4D[
  subDDistinct,
  <|1 -> -1, 2 -> 1, 3 -> 1, P -> -1|>,
  ParentLeg -> P, ReferenceLeg -> r,
  PolarizationReferences -> <|P -> eta[P]|>,
  PrimeParentHelicity -> 1,
  PrimePolarizationReference -> etaPrime[P],
  MomentumFraction -> x];
\end{verbatim}
The last two options independently specify the helicity and reference spinor of \texttt{ePrime[P]}.  Changing either parent reference changes only the gauge representative and leaves the reduced result invariant.
\end{samepage}
The accompanying file \texttt{demo.nb} collects these commands and further examples in a single executable session.  


\section{Conclusions and outlooks}
\label{sec:conclusions}

We have studied two recursive descriptions of the strict subleading adjacent-collinear limit of tree-level Yang--Mills amplitudes.  The general-dimensional construction uses a Lorentz null rotation to transport physical polarizations along the collinear path.  It separates the finite coefficient into a channel-deleted hard term and a factorization contribution.  The modified BG recursion evaluates the former directly on the collinear surface, while the latter can be obtained by ordinary BG recursion. In four dimensions, the same coefficient follows from a collinear-aware BCFW recursion.  Terms in which the collinear pair is separated are evaluated directly on the collinear surface, whereas terms in which the pair remains together are treated recursively.  MHV and $\overline{\rm MHV}$ amplitudes provide closed terminal data.  This organization avoids a Laurent expansion of the complete amplitude and applies at arbitrary momentum fraction $x$.

The construction also accommodates two distinct parent polarizations.  At the symmetric split, the equal-polarization result is manifestly independent of the auxiliary collinear data.  For distinct polarizations, the explicit factorization contribution is antisymmetric under exchange of the two parent states and therefore cancels in the polarization-symmetrized combination. Its four-dimensional reduction reproduces the corresponding cancellation between the ordered mixed-helicity limits. Finally, the exact open--closed disk relation turns the recursive Yang--Mills data into a representation of one-graviton Einstein--Yang--Mills amplitudes and their higher-derivative corrections. A self-contained \textsc{Mathematica} implementation is provided as ancillary material accompanying the paper. Our work has suggested several future directions:

\paragraph{Loop-level collinear limits.}
A natural next step is to extend the construction beyond tree level. Universal leading collinear factorization is well understood at one loop and beyond: an $L$-loop amplitude receives contributions from an $\ell$-loop splitting amplitude multiplying an $(L-\ell)$-loop lower-point amplitude~\cite{Bern:1994zx,Kosower:1999rx,Kosower:1999xi}. Considerably less is known finite subleading term. In dimensional regularization, factors such as $s_{n-1,n}^{-\epsilon_{\rm IR}}$ generate logarithms of the collinear parameter, so the analogue of the strict coefficient must be defined either before integration or for an infrared-subtracted finite remainder. It would be interesting to formulate a loop-level channel-deleted recursion or generalized-unitarity construction and to determine which part of the subleading limit remains universal after infrared subtraction. The covariant null-rotation path used here supplies a fixed continuation of both momenta and polarizations and should therefore provide a useful starting point for separating genuine subleading data from regulator-dependent logarithms. Moreover, it would be interesting to investigate the collinear limit at the integrand level, particularly within the framework of surface ``The'' Yang--Mills integrand, which is uniquely defined and exhibits exact loop cuts and factorization properties~\cite{Arkani-Hamed:2023swr,Arkani-Hamed:2023jry,Arkani-Hamed:2024tzl}.

\paragraph{Collinear limits in gravity.}
A second direction is the corresponding finite collinear limit of gravitational amplitudes. At leading order, gravitational splitting amplitudes are naturally related to products of gauge-theory splitting amplitudes through Kawai--Lewellen--Tye or Bern-Carrasco-Johansson double-copy relations~\cite{Kawai:1985xq,Bern:1998sv,Bern:2008qj, Bern:2010ue,Bern:2019prr}. These relations suggest applying the double copy directly to the recursive Yang--Mills construction developed here. At subleading order, however, one must also expand the momentum kernel and combine singular and regular terms from the two gauge-theory copies, so the answer need not follow from a naive product of subleading splitting functions. A complementary approach would be to construct a direct gravitational BG or BCFW recursion along the same null-rotation path. Comparing the direct and double-copy constructions would clarify how the hard contribution, contact terms, and dilaton or antisymmetric-tensor states are organized when color ordering and adjacency are absent.

\paragraph{Celestial OPEs.}
Momentum-space collinear limits become operator-product expansions (OPEs) after Mellin transformation to the celestial basis. The leading splitting functions determine the singular celestial OPE coefficients of gluons and gravitons~\cite{Fan:2019emx,PateRaclariuStromingerYuan2019}, while worldsheet and on-shell recursion methods have exposed descendant contributions and, in special sectors, the OPE to all orders in the celestial separation~\cite{AdamoBuCasaliSharma2021,AdamoBuCasaliSharma2022,RenSchreiberSharmaWang2023}. The strict finite coefficients computed here should provide systematic momentum-space input for the first terms beyond the leading collinear singularity. At loop level this question is intertwined with infrared subtraction: collinear logarithms lead to derivatives with respect to celestial conformal dimensions and to logarithmic operator structures~\cite{Bhardwaj:2022anh}. Understanding this map may provide a direct bridge between recursive amplitude methods and the descendant structure, loop corrections, and consistency conditions of celestial OPEs.

\acknowledgments
We would like to thank Lance Dixon for useful discussions. This work is supported by the DFG grant 508889767, Forschungsgruppe ``Modern foundations of scattering amplitudes''.

\newpage
\appendix

\section{Four-dimensional reduction of the covariant collinear limits}
\label{app:four-dimensional-reduction}

We verify here that the covariant leading collinear formulas reproduce the standard four-dimensional splitting amplitudes for both equal and opposite helicities. The spinor parametrization \eqref{eq:4d-collinear-parametrization} implies
\begin{equation}
 \langle n-1,n\rangle=\epsilon\langle Pr\rangle,
 \qquad
 [n-1,n]=\epsilon[Pr].
\label{eq:app-spinor-pair}
\end{equation}
Recalling $c=\sqrt{x}$ and $s=\sqrt{1-x}$, the corresponding transverse vector in Eq.~\eqref{eq: 4d kperp} obeys
\begin{equation}
 k_\perp^2=-2c^2s^2\langle Pr\rangle[Pr]\,.
\label{eq:app-kperp-square}
\end{equation}
Together with $K=r$, this is equivalently $k_\perp^2=-2x(1-x)P\cdot K$ in the vector normalization used throughout the paper.  It ensures that the covariant momentum parametrization reproduces Eq.~\eqref{eq:4d-momentum-expansion}, including its $O(\epsilon^2)$ terms.

We represent the two helicity polarizations of the parent momentum by
\begin{equation}
 \varepsilon_P^{+\mu}(q_+)
 =-\frac{1}{\sqrt{2}}\frac{\langle q_+|\gamma^\mu|P]}{\langle Pq_+\rangle},
 \qquad
 \varepsilon_P^{-\mu}(q_-)
 =-\frac{1}{\sqrt{2}}\frac{\langle P|\gamma^\mu|q_-]}{[Pq_-]}.
\label{eq:app-helicity-polarizations}
\end{equation}
We use the standard gamma-matrix identity
\begin{equation*}
 \langle a|\gamma^\mu|b]\langle c|\gamma_\mu|d]
 =2\langle ac\rangle[bd].
\end{equation*}
Equivalently, the vector associated with a rank-one bispinor is $(|a\rangle[b|)^\mu:=\langle a|\gamma^\mu|b]/\sqrt{2}$, for which $(|a\rangle[b|)\cdot(|c\rangle[d|)=\langle ac\rangle[bd]$.  The factors $1/\sqrt{2}$ in Eq.~\eqref{eq:app-helicity-polarizations} therefore give the physical normalization
\begin{equation}
 (\varepsilon_P^+)^2=(\varepsilon_P^-)^2=0,
 \qquad
 \varepsilon_P^+\cdot\varepsilon_P^-=-1.
\label{eq:app-helicity-normalization}
\end{equation}
For example,
\begin{equation}
 \varepsilon_P^+(q_+)\cdot\varepsilon_P^-(q_-)
 =\frac{1}{2}\frac{2\langle q_+P\rangle[Pq_-]}
 {\langle Pq_+\rangle[Pq_-]}=-1.
\end{equation}
The result is independent of $q_+$ and $q_-$, as a change of either reference adds a multiple of $P^\mu$.  Direct contraction with $k_\perp$ gives
\begin{equation}
 \varepsilon_P^+\cdot k_\perp=cs[Pr],
 \qquad
 \varepsilon_P^-\cdot k_\perp=cs\langle Pr\rangle.
\label{eq:app-pol-kperp}
\end{equation}

\paragraph{Equal helicities.}
For $\varepsilon'_P=\varepsilon_P$, the covariant leading result is
\begin{equation}
 \left.A_n(\varepsilon_P,\varepsilon_P)\right|_{\rm coll.}^{(-1)}
 =-\frac{2\,\varepsilon_P\cdot k_\perp}
 {\epsilon k_\perp^2}
 A_{n-1}(1,2,\ldots,n-2,P).
\label{eq:app-equal-D-leading}
\end{equation}
Equations~\eqref{eq:app-kperp-square} and \eqref{eq:app-pol-kperp} therefore give
\begin{equation}
\begin{aligned}
 \left.A_n(\ldots,(n-1)^+,n^+)\right|_{\rm coll.}^{(-1)}
 &=\frac{1}{\epsilon cs\langle Pr\rangle}
 A_{n-1}(\ldots,P^+),
 \\
 \left.A_n(\ldots,(n-1)^-,n^-)\right|_{\rm coll.}^{(-1)}
 &=\frac{1}{\epsilon cs[Pr]}
 A_{n-1}(\ldots,P^-).
\end{aligned}
\label{eq:app-equal-4d-leading}
\end{equation}
Using Eq.~\eqref{eq:app-spinor-pair}, these are the familiar same-helicity splitting factors $1/(cs\langle n-1,n\rangle)$ and $1/(cs[n-1,n])$, in the orientation and normalization used in this paper.

\paragraph{Opposite helicities.}
The term proportional to $k_\perp^\mu$ in Eq.~\eqref{eq:mixed-leading-D} does not vanish for opposite helicities. Rather, it is required to obtain the correct four-dimensional weights.  Set
\begin{equation}
 \varepsilon_P=\varepsilon_P^- ,
 \qquad
 \varepsilon'_P=\varepsilon_P^+,
 \qquad
 A(P^\pm):=\varepsilon_P^{\pm\mu}
 A_\mu(1,2,\ldots,n-2,P).
\label{eq:app-lower-helicity-amplitudes}
\end{equation}
The Ward identity $P^\mu A_\mu=0$ and transverse completeness imply Eq.~\eqref{eq:app-kperp-current-decomposition}.  To see this explicitly, one may choose a common null reference vector $Q$ for the two helicity polarizations; using different reference spinors changes them only by terms proportional to $P^\mu$.  With the normalization in Eq.~\eqref{eq:app-helicity-normalization}, completeness reads
\begin{equation*}
 \eta^{\mu\nu}
 =-\varepsilon_P^{+\mu}\varepsilon_P^{-\nu}
  -\varepsilon_P^{-\mu}\varepsilon_P^{+\nu}
  +\frac{P^\mu Q^\nu+Q^\mu P^\nu}{P\cdot Q}.
\end{equation*}
Contracting this identity with $k_{\perp,\mu}A_\nu$ gives
\begin{equation*}
 \begin{aligned}
 k_\perp^\mu A_\mu={}&
 -(k_\perp\cdot\varepsilon_P^+)(\varepsilon_P^-\cdot A)
 -(k_\perp\cdot\varepsilon_P^-)(\varepsilon_P^+\cdot A)
 \\
 &+\frac{(k_\perp\cdot P)(Q\cdot A)
 +(k_\perp\cdot Q)(P\cdot A)}{P\cdot Q}.
 \end{aligned}
\end{equation*}
The last line vanishes because $k_\perp\cdot P=0$ and $P\cdot A=0$.  Using the definition of $A(P^\pm)$ in Eq.~\eqref{eq:app-lower-helicity-amplitudes} then yields
\begin{equation}
 k_\perp^\mu A_\mu
 =-(k_\perp\cdot\varepsilon_P^+)A(P^-)
  -(k_\perp\cdot\varepsilon_P^-)A(P^+).
\label{eq:app-kperp-current-decomposition}
\end{equation}
Longitudinal terms in the complete polarization sum drop out because both $P\cdot k_\perp$ and $P^\mu A_\mu$ vanish.  The vector contraction in Eq.~\eqref{eq:mixed-leading-D} consequently reduces to
\begin{equation}
\begin{aligned}
 &x(\varepsilon_P^+\cdot k_\perp)A(P^-)
 +(1-x)(\varepsilon_P^-\cdot k_\perp)A(P^+)
 \\
 &\quad
 -x(1-x)(\varepsilon_P^-\cdot\varepsilon_P^+)
 k_\perp^\mu A_\mu
 \\
 &=x^2(\varepsilon_P^+\cdot k_\perp)A(P^-)
 +(1-x)^2(\varepsilon_P^-\cdot k_\perp)A(P^+).
\end{aligned}
\label{eq:app-mixed-reduction}
\end{equation}
Thus $k_\perp^\mu A_\mu$ is not an additional lower-point object: in four dimensions it is a linear combination of the two ordinary helicity amplitudes.  Substitution of Eqs.~\eqref{eq:app-kperp-square} and \eqref{eq:app-pol-kperp} gives
\begin{equation}
\begin{aligned}
 \left.A_n(\ldots,(n-1)^-,n^+)\right|_{\rm coll.}^{(-1)}
 ={}&\frac{c^3}{s}\frac{1}{\epsilon\langle Pr\rangle}A(P^-)
 +\frac{s^3}{c}\frac{1}{\epsilon[Pr]}A(P^+),
 \\
 \left.A_n(\ldots,(n-1)^+,n^-)\right|_{\rm coll.}^{(-1)}
 ={}&\frac{s^3}{c}\frac{1}{\epsilon\langle Pr\rangle}A(P^-)
 +\frac{c^3}{s}\frac{1}{\epsilon[Pr]}A(P^+).
\end{aligned}
\label{eq:app-mixed-4d-leading}
\end{equation}
These are precisely the standard mixed-helicity splitting amplitudes.  In an MHV configuration with one negative-helicity hard leg, $A(P^+)=0$; the first line of Eq.~\eqref{eq:app-mixed-4d-leading} then reduces to the leading term in Eq.~\eqref{eq:mixed-MHV-expansion}.  The $k_\perp^\mu A_\mu$ term in the covariant formula is essential: without it the powers $c^3/s$ and $s^3/c$ would be replaced by incorrect momentum-fraction weights.

The same reduction applies to the finite coefficients.  The modified Berends--Giele hard contribution is built entirely from Lorentz contractions, so its restriction to four dimensions commutes with the strict collinear replacement.  The remaining explicit factorization terms in Eqs.~\eqref{eq: subleading} and~\eqref{eq:mixed-subleading-D} reduce under the same rules to the finite same- and mixed-helicity terms used in the BCFW construction.  In particular, the equal-helicity pole correction vanishes at $x=1/2$, while the two ordered mixed-helicity corrections are exchanged by $\varepsilon_P^-\leftrightarrow\varepsilon_P^+$ and their reference dependence cancels in the polarization-symmetrized sum.

\section{Consistency of the modified Berends--Giele recursion}
\label{app:modified-bg}

Section~\ref{subsec:modified-bg} defines the channel-deleted current and the associated amplitude $A_n^{\rm mod}$. We now prove that its strict value equals the pole-subtracted hard contribution. The point requiring proof is that setting $J^\mu(n-1,n)$ to zero might appear to discard a finite term when a numerator factor cancels its propagator $1/s_{n-1,n}$.

The special current $J^\mu(n-1,n)$ occurs at most once in any tree graph.  All terms containing it may therefore be collected into a complete amputated hard-side current $\mathcal J_\mu(Q)$, where $Q:=p_{n-1}+p_n$.  Concretely, $\mathcal J_\mu(Q)$ is the coefficient obtained by replacing the special subcurrent by an arbitrary vector and summing all BG graphs above it.  It is therefore a complete current, rather than a graph-by-graph object; its dependence on $Q$ enters through the hard-side vertices and propagators.  Before taking the collinear limit, the amplitude separates exactly as
\begin{equation}
 A_n(\epsilon)
 =A_n^{\rm mod}(\epsilon)
 +\frac{N_{n-1,n}^{\mu}(\epsilon)\,
 \mathcal J_\mu(Q(\epsilon))}
 {s_{n-1,n}(\epsilon)},
 \qquad
 N_{n-1,n}^{\mu}:=s_{n-1,n}J^\mu(n-1,n).
\label{eq:app-bg-channel-split}
\end{equation}
At $Q^2=0$, the current reduces to the polarization-stripped lower-point amplitude,
\begin{equation*}
 \mathcal J_\mu(P)=A_\mu(1,2,\ldots,n-2,P).
\end{equation*}
Since all its other external legs are on shell, the complete BG current also obeys the off-shell Ward identity
\begin{equation}
 Q^\mu\mathcal J_\mu(Q)=0.
\label{eq:app-bg-off-shell-Ward}
\end{equation}
This identity follows inductively from the ordinary BG recursion: contractions of the cubic vertices with $Q^\mu$ cancel among adjacent deconcatenations and against the quartic-vertex terms.  It holds only after the complete hard-side current has been assembled.

The only possible concern with setting $J^\mu(n-1,n)$ to zero is that a factor of $s_{n-1,n}$ in the numerator of the second term in Eq.~\eqref{eq:app-bg-channel-split} may cancel the pair propagator and leave a finite contribution.  To determine where this contribution belongs, note that the collinear path gives the exact relation
\begin{equation}
 Q^\mu=P^\mu+\frac{s_{n-1,n}}{2P\cdot K}K^\mu.
\label{eq:app-bg-Q-path}
\end{equation}
A convenient conserved continuation of the on-shell hard current is
\begin{equation}
 \widehat{\mathcal J}_\mu(Q)
 :=\mathcal J_\mu(P)
 -\frac{s_{n-1,n}}{2(P\cdot K)^2}
 \bigl[K\cdot\mathcal J(P)\bigr]K_\mu,
 \qquad
 Q^\mu\widehat{\mathcal J}_\mu(Q)=0.
\label{eq:app-bg-conserved-continuation}
\end{equation}
At generic hard kinematics, both the exact current and this continuation are analytic near $s_{n-1,n}=0$ and agree there.  Consequently,
\begin{equation}
 \mathcal J_\mu(Q)-\widehat{\mathcal J}_\mu(Q)
 =s_{n-1,n}\,\Delta_\mu(Q),
 \qquad
 P^\mu\Delta_\mu(P)=0,
\label{eq:app-bg-current-difference}
\end{equation}
where the second relation follows by subtracting their Ward identities and then taking $Q\to P$.

It remains to use the special form of the pair numerator.  Including all three cubic tensor structures, it is
\begin{equation}
 N_{n-1,n}^{\mu}
 =-2\varepsilon_{n-1}^\mu
   (p_{n-1}\cdot\varepsilon_n)
 +2\varepsilon_n^\mu
   (p_n\cdot\varepsilon_{n-1})
 +(\varepsilon_{n-1}\cdot\varepsilon_n)
 (p_{n-1}-p_n)^\mu.
\label{eq:app-bg-pair-numerator}
\end{equation}
For two parent polarizations $\varepsilon_P$ and $\varepsilon'_P$, its strict value is
\begin{equation}
 N_{n-1,n}^{\mu}\big|_{\epsilon=0}
 =(2x-1)(\varepsilon_P\cdot\varepsilon'_P)P^\mu.
\label{eq:app-bg-full-numerator}
\end{equation}
For equal polarizations this vanishes because $\varepsilon_P^2=0$.  For distinct polarizations it need not vanish, but it is purely longitudinal.  Equations~\eqref{eq:app-bg-current-difference} and~\eqref{eq:app-bg-full-numerator} now imply
\begin{equation}
 \left.
 \frac{N_{n-1,n}^{\mu}
 \bigl[\mathcal J_\mu(Q)-\widehat{\mathcal J}_\mu(Q)\bigr]}
 {s_{n-1,n}}
 \right|_{\epsilon^0}
 =(2x-1)(\varepsilon_P\cdot\varepsilon'_P)
 P^\mu\Delta_\mu(P)=0.
\label{eq:app-bg-no-missing-hard-term}
\end{equation}
To see explicitly how this implies the desired result, subtract the continued factorization block from the exact decomposition \eqref{eq:app-bg-channel-split}:
\begin{equation*}
 A_n-\frac{N_{n-1,n}^{\mu}\widehat{\mathcal J}_\mu(Q)}
 {s_{n-1,n}}
 =A_n^{\rm mod}
 +\frac{N_{n-1,n}^{\mu}
 \bigl[\mathcal J_\mu(Q)-\widehat{\mathcal J}_\mu(Q)\bigr]}
 {s_{n-1,n}}.
\end{equation*}
The modified amplitude contains no $(n{-}1,n)$ propagator and is therefore regular at $\epsilon=0$; its finite coefficient is simply its strict value.  Taking the $\epsilon^0$ coefficient of this identity, the second term on the right-hand side vanishes by Eq.~\eqref{eq:app-bg-no-missing-hard-term}.  Hence
\begin{equation}
 \left.
 \left[A_n-\frac{N_{n-1,n}^{\mu}
 \widehat{\mathcal J}_\mu(Q)}{s_{n-1,n}}
 \right]\right|_{\epsilon^0}
 =\left.A_n^{\rm mod}\right|_{\epsilon=0}.
\label{eq:app-bg-hard-equality}
\end{equation}
The left-hand side of Eq.~\eqref{eq:app-bg-hard-equality} is precisely the pole-subtracted amplitude evaluated in the strict collinear limit.  Indeed, $\widehat{\mathcal J}_\mu(P)=\mathcal J_\mu(P)=A_\mu(1,2,\ldots,n-2,P)$, so $N_{n-1,n}^{\mu}\widehat{\mathcal J}_\mu(Q)$ is a continuation of the on-shell factorization residue away from $s_{n-1,n}=0$.  The defining condition $Q\cdot\widehat{\mathcal J}(Q)=0$ makes this continuation compatible with the Ward identity.  Equation~\eqref{eq:app-bg-no-missing-hard-term} then shows that replacing the exact hard-side current by this conserved continuation does not change the strict finite coefficient. This proves that the modified recursion computes the strict hard contribution for both equal and distinct polarizations.

\newpage
\bibliographystyle{JHEP}
\bibliography{Refs}
\end{document}